\documentclass{aa}  

\usepackage{amstext,amsmath}
\usepackage{graphicx}
\usepackage{textcomp}
\usepackage{txfonts}
\usepackage{time}
\usepackage{subfig}
\usepackage{multicol}
\usepackage{color}
\usepackage{url}
\usepackage{pstricks} 
\usepackage{natbib,twoopt}
\usepackage{bm}
\usepackage{gensymb}
\usepackage[flushleft]{threeparttable}
\bibpunct{(}{)}{;}{a}{}{,} 
\makeatletter
\newcommandtwoopt{\citeads}[3][][]{\href{http://adsabs.harvard.edu/abs/#3}%
{\def\hyper@linkstart##1##2{}%
\let\hyper@linkend\@empty\cite[#1][#2]{#3}}}
\newcommandtwoopt{\citealtads}[3][][]{\href{http://adsabs.harvard.edu/abs/#3}%
{\def\hyper@linkstart##1##2{}%
\let\hyper@linkend\@empty\citealt[#1][#2]{#3}}}
\newcommandtwoopt{\citepads}[3][][]{\href{http://adsabs.harvard.edu/abs/#3}%
{\def\hyper@linkstart##1##2{}%
\let\hyper@linkend\@empty\citep[#1][#2]{#3}}}
\newcommandtwoopt{\citetads}[3][][]{\href{http://adsabs.harvard.edu/abs/#3}%
{\def\hyper@linkstart##1##2{}%
\let\hyper@linkend\@empty\citet[#1][#2]{#3}}}
\newcommandtwoopt{\citeyearads}[3][][]%
{\href{http://adsabs.harvard.edu/abs/#3}
{\def\hyper@linkstart##1##2{}%
\let\hyper@linkend\@empty\citeyear[#1][#2]{#3}}}
\makeatother

\newcommand{\hed}{He~\textsc{i} D\textsubscript{3}\,}
\newcommand{\hei}{He~\textsc{i} 10830\,}
\newcommand{\der}{{\rm d}}

\usepackage{fixltx2e}
\usepackage{url}
\usepackage{epic}
\usepackage{rotating}
\usepackage{breqn}
\usepackage{multirow} 
\usepackage{balance}
\usepackage{epic}
\usepackage{xcolor}
\definecolor{dark-gray}{gray}{0.4}
\usepackage[colorlinks=true,urlcolor=blue,citecolor=dark-gray,linkcolor=blue]{hyperref}

\usepackage{xifthen}

\makeatletter
\def\url@leostyle{%
  \@ifundefined{selectfont}{\def\UrlFont{\sf}}{\def\UrlFont{\tiny\ttfamily}}}
\makeatother
\begin{document}
\authorrunning{T. Libbrecht et al. }

\title{Chromospheric condensations and magnetic field in a C3.6-class flare studied via \hed spectro-polarimetry}

\author{Tine Libbrecht\inst{1}, Jaime de la Cruz Rodr\'iguez\inst{1}, Sanja Danilovic\inst{1}, Jorrit Leenaarts\inst{1}, Hiva Pazira\inst{1}} 
\institute{Institute for Solar Physics, Dept. of Astronomy, Stockholm University, Albanova University Center, SE-10691 Stockholm, Sweden}

\frenchspacing
\abstract
{Magnetic reconnection during flares takes place in the corona, but a substantial part of flare energy is deposited in the chromosphere. However, high-resolution spectro-polarimetric chromospheric observations of flares are very rare. The most used observables are Ca~\textsc{ii} 8542 \AA\;and \hei \AA.}
{We aim to study the chromosphere during a C3.6 class flare via spectro-polarimetric observations of the \hed line.}
{We present the first SST/CRISP spectro-polarimetric observations of \hed. We analyzed the data using the inversion code \textsc{Hazel}, and estimate the LOS velocity and the magnetic field vector.}
{Strong \hed emission at the flare footpoints, as well as strong \hed absorption profiles tracing the flaring loops are observed during the flare. The \hed traveling emission kernels at the flare footpoints exhibit strong chromospheric condensations of up to $\sim 60\,\rm km~s^{-1}$ at their leading edge. Our observations suggest that such condensations result in shocking the deep chromosphere, causing broad and modestly blueshifted \hed profiles indicating subsequent upflows. A strong and rather vertical magnetic field of up to $\sim 2500$ G is measured in the flare footpoints, confirming that the \hed line is likely formed in the deep chromosphere at those locations. We provide chromospheric line-of-sight velocity and magnetic field maps obtained via \hed inversions. We propose a fan-spine configuration as the flare magnetic field topology.}
{The \hed line is an excellent diagnostic to study the chromosphere during flares. The impact of strong condensations on the deep chromosphere has been observed. Detailed maps of the flare dynamics and the magnetic field are obtained.}


\maketitle

\section{Introduction}\label{sec:intro}
Both observations and models agree that the chromosphere is greatly affected by solar flares, even though the magnetic reconnection happens in the corona. The sites of energy deposition in the chromosphere during flares manifest themselves as bright ribbons in H$\alpha$ and other chromospheric diagnostics such as Ca \textsc{ii} 8542 \AA, and Mg \textsc{ii} h\&k (e.g.,~\citealtads{1966SSRv....5..388S,1972SoPh...24..414H} or reviews by \citealtads{lrsp-2008-1,2011SSRv..159...19F}).

The appearance of flares in neutral helium triplet lines \hed at 5876 \AA, and \hei is slightly different: strong emission kernels are often observed during flares in those lines (for a clear example, see e.g., \citealtads{2014ApJ...793...87Z}). Those emission kernels are usually located within the flaring ribbons, likely tracing the flare footpoints that are the sites of hard X-rays (HXRs) originating from non-thermal electrons impacting the chromosphere \citepads{1980ApJ...235..618Z,1981ApJ...248L..45Z,1983ApJ...271..832F,2013ApJ...774...60L,1990ApJS...73..111Z}.

For this paper, we observed both bright \hed emission kernels as well as a large \hed absorption structure in a C3.6 class flare. Many observations are available, showing that both emission and absorption can be present during flares in neutral helium lines (e.g.,~\citealtads{2011A&A...526A..42S,2014ApJ...793...87Z,2016ApJ...819...89X}). The line formation mechanism for \hed and He \textsc{i} 10830 \AA\;in flares is debated in many observational studies, but very few modeling results are available. The photoionization-recombination mechanism (PRM) is the established line formation mechanism for He \textsc{i} 10830 \AA\;and \hed in the quiet sun and possibly in active regions \citepads{1975ApJ...199L..63Z,2005ApJ...619..604M,2008ApJ...677..742C,2016A&A...594A.104L}. However, in flare footpoints, heating via thermal conduction or collisions with non-thermal electrons might be dominant \citepads{2005A&A...432..699D}.

Only in the last one to one-and-a-half decades, advances in ground-based chromospheric observations and mostly in non-LTE inversion tools made it possible to measure the chromospheric magnetic field, even though signal-to-noise ratio in the Stokes parameters continues to pose challenges \citepads{2017SSRv..210..109D}. In addition, obtaining ground-based chromospheric flare observations requires a decent portion of persistence and luck. This results in the very limited number of studies employing high spatial resolution spectro-polarimetry to study the chromosphere during flares \citepads{2014A&A...561A..98S,2015IAUS..305...73K,2015ApJ...814..100J,2017ApJ...846....9K,2017ApJ...834...26K,2018arXiv180500487K}. Inverting spectro-polarimetric data provides magnetic field and velocity maps, which are invaluable to study flare impact on the chromosphere. They could also serve as input for magnetic field extrapolations in flares or to cross-validate output results of extrapolations. Therefore, this study aims to obtain chromospheric line-of-sight velocity and magnetic field maps during the observed C3.6-class flare via inversions of the \hed line. Thus far, all chromospheric flare inversions have been conducted via spectro-polarimetry of Ca \textsc{ii} 8542 \AA, or He \textsc{i} 10830 \AA. This paper adds another chromospheric diagnostic to this set to be used in flares. 

Previous results obtained by the observations of \hed line in flares consist of imaging data obtained at BBSO or NSO/Sac Peak \citepads{1980ApJ...235..618Z,1981ApJ...248L..45Z,1990ApJS...73..111Z,1992A&A...256..255F,1996A&A...306..625C,2013ApJ...774...60L}, but those imaging data have limited spatial resolution. Flaring spectra of \hed have been used in the context of measuring chromospheric condensations \citepads{1995PASJ...47..239S,2003MmSAI..74..635T,2006SoPh..239..193F} and helium abundance measurement \citepads{2008ApJ...681..650A}.

In this paper, we show that \hed is an outstanding diagnostic to measure chromospheric velocities and to study chromospheric condensation (downflows) and evaporation (upflows). This flare phenomenon has received increased attention since the launch of the Interface Region Imaging Spectrograph (IRIS, \citealtads{2014SoPh..289.2733D}). Chromospheric condensations have been studied using Mg \textsc{ii} h\&k and the subordinate triplet, C \textsc{ii} 1330 \AA\,and/or Si \textsc{iv} 1400 \AA\,lines \citepads{2014ApJ...797L..14T,2015ApJ...811..139T,2015ApJ...811....7L,2015ApJ...807L..22G,2016ApJ...816...89P,2017ApJ...848..118L,2017ApJ...848...39B,2018PASJ..tmp...61T}. Mostly strong downflows have been measured with strengths varying between $10 - 80\,\rm km~s^{-1}$. Blue asymmetries and upflows in Mg \textsc{ii} h\&k have been reported in certain cases \citepads{2015A&A...582A..50K,2018PASJ..tmp...61T}, reminiscent of blue asymmetries reported in H$\alpha$ and Ca~\textsc{ii} 8542 \AA~\citepads{1990ApJ...363..318C,1994SoPh..152..393H,2015ApJ...813..125K}. 

The difficulty with measuring upflows in H$\alpha$, Ca~\textsc{ii} 8542 \AA, and  Mg \textsc{ii} h\&k is that these lines are optically thick, formed in an extended range of photospheric to chromospheric heights and often exhibit self-reversal in the line core. These properties pose challenges for interpreting and modeling of the lines: blue asymmetries can be either due to a velocity gradient including upflows, or a redshift in the layer where the line core is formed, so that the blue wing is enhanced. This type of ambiguity is abscent in the \hed line, which is generally formed in a thin layer in the upper chromosphere and usually (but not always) either in complete emission or absorption. In those cases, the interpretation of the velocity is very straightforward, where redshift simply corresponds to downflows and blueshift to upflows. 

In recent years, time-dependent 1D-modeling of the flaring atmosphere has been conducted with use of the \textsc{RADYN} code and chromospheric radiative diagnostics can be synthesized (e.g. \citealtads{2005ApJ...630..573A,2015ApJ...809..104A,2016ApJ...827..101K,2016ApJ...827...38R,2017ApJ...836...12K,2017ApJ...842...82R}). We suggest that synthesizing \hed would be a great addition to constrain the flaring atmosphere in such models. The \hei line is by default synthesized in RADYN \citepads{2015ApJ...809..104A} but no in-depth studies of the line with RADYN are currently available.

To summarize, this paper focuses on obtaining chromospheric magnetic field and velocity maps, and examines chromospheric condensations in more detail. The outline of the paper is the following: Sect.~\ref{sec:obs} describes the flare and the observations with SST/CRISP, Sect.~\ref{sec:ana} gives a detailed description of the classification and inversion methods used in this paper, Sect.~\ref{sec:res} describes the main results, followed by a discussion in Sect.~\ref{sec:disc} where the results are put in context of the current literature. A summary and conclusions are given in Sect.~\ref{sec:concl}.

\section{Observations and data overview}\label{sec:obs}
\subsection{Event overview}\label{subs:descr}
\begin{figure}
\center
\includegraphics[scale=1]{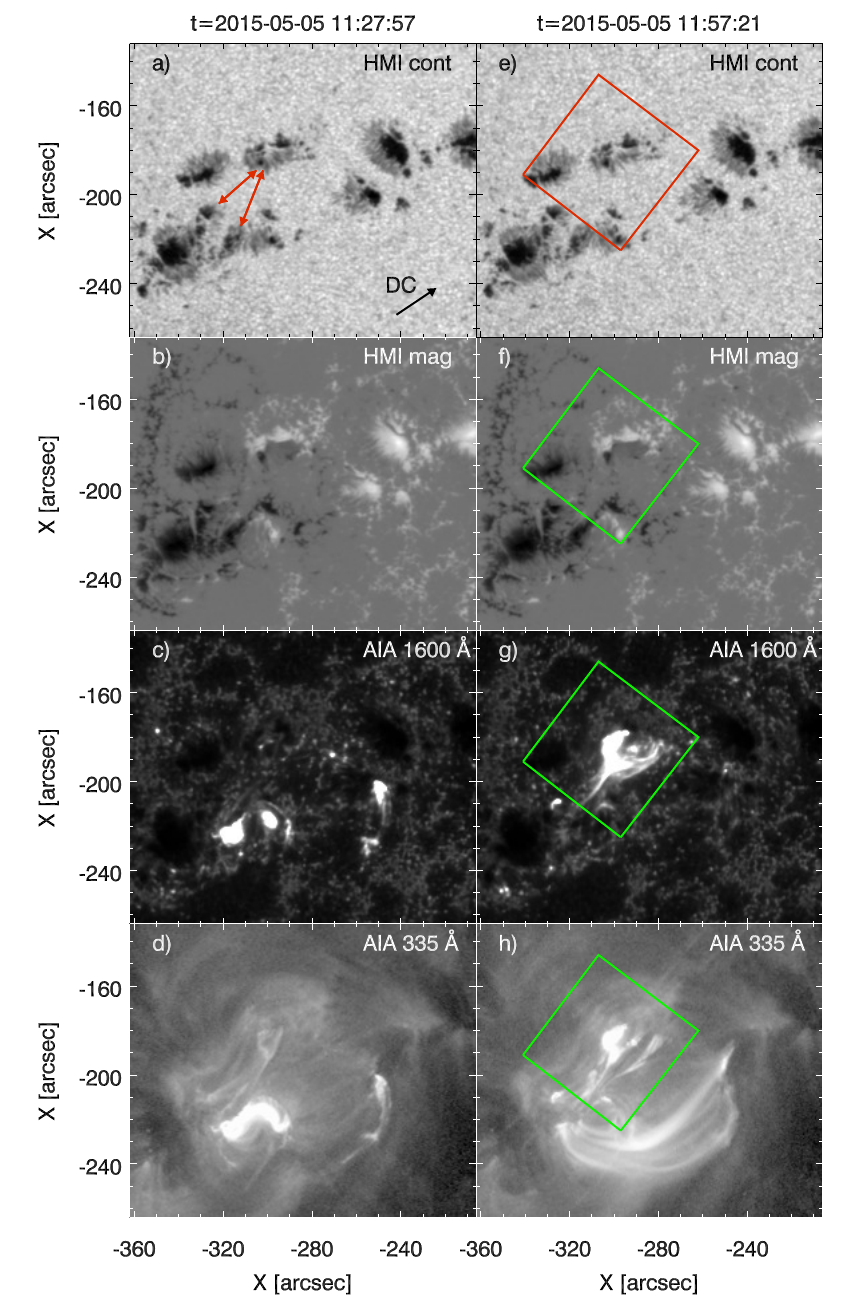}
\caption{Overview of AR 12335. Left column: a C6.6 class flare preceding the flare studied in this paper, shown at its peak time. Right column: the C3.6 class flare, shown at its peak time. The SST field of view is given in red and green squares. The direction of disk center (DC) is indicated in panel a) with a black arrow. The red arrows in panel a) show regions of previous flux emergence in the AR. The data is taken from SDO/HMI and SDO/AIA, with the wavelength band shown in the upper right corner of each panel. \label{fig:fieldov_mag}}
\end{figure}

The current work focuses on a C3.6 GOES-class flare which occurred on 2015-05-05 in AR 12335. This active region turned around the east limb on 2015-04-30 containing several small sunspots and pores. In the subsequent days, AR 12335 grew in size, number of sunspots and complexity. Fig.~\ref{fig:fieldov_mag} shows AR 12335 at the time of two flares: a C6.6 GOES-class flare and the subject of this study: a C3.6 class flare. The red arrows in panel a) of Fig.~\ref{fig:fieldov_mag} indicate areas where substantial flux emergence on 2015-05-03 and 2015-05-04 has formed developing and unorganized sunspots. AR 12335 showed high levels of flaring activity between 2015-04-30 and 2015-05-07.

\begin{figure}
\center
\includegraphics[scale=1]{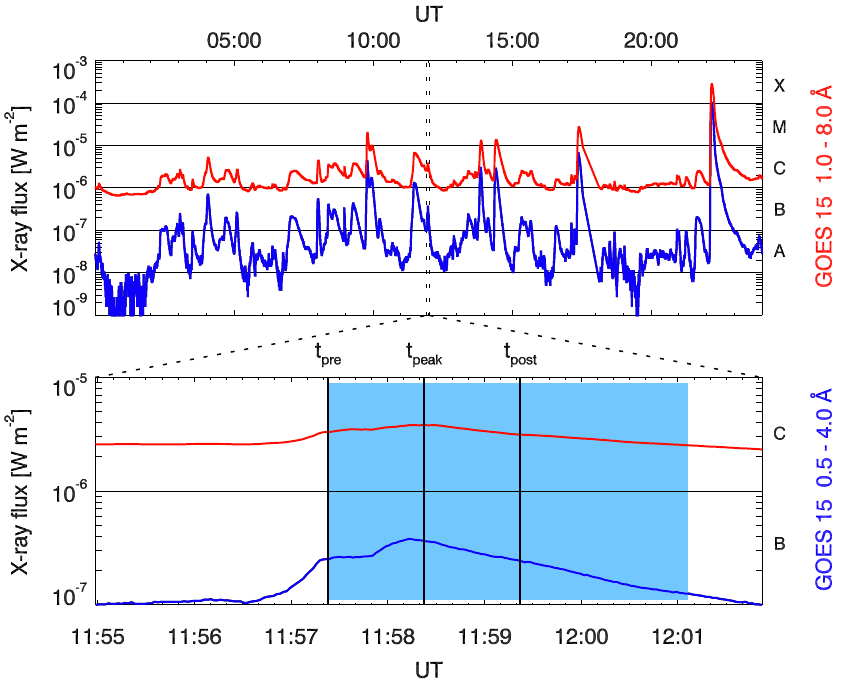}
\caption{Time evolution of the X-ray flux as measured by the GOES satellite in channel $1.0 - 8.0$ \AA\;(red) and $0.5 - 4.0$ \AA\;(blue). The upper panel shows the X-ray flux for the entire day 2015-05-05. The lower panel focusses on the C3.6 class flare in AR 12335. The blue area indicates the time of SST observation. The black vertical lines show the times of the three scans that we have used in our inversions, referred to as $t_{\rm pre}$, $t_{\rm peak}$ and $t_{\rm post}$. \label{fig:goes}}
\end{figure}

Figure~\ref{fig:goes} displays GOES light curves for 2015-05-05. The lower panel of Fig.~\ref{fig:goes} zooms in on the C3.6 GOES-class flare at coordinates $x=-300\arcsec,y=-190\arcsec$ with a heliocentric viewing angle of $\mu=0.93$. The flare started at 11:55 and lasted for five minutes with a peak time at 11:58\footnote{According to \url{http://www.lmsal.com/solarsoft/latest_events_archive/events_summary/2015/05/05/gev_20150505_1155/index.html}\label{note1}}. 

\subsection{SST observation}\label{subs:sst}
The lower panel of Fig.~\ref{fig:goes} indicates the time interval for which we have obtained data from the Swedish 1-m Solar Telescope (SST, \citealtads{2003SPIE.4853..341S}) on La Palma, employing the CRisp Imaging SpectroPolarimeter (CRISP, \citealtads{2008ApJ...689L..69S}). The observing sequence consisted of 16 wavelength positions sampling the \hed line with polarimetry: $\lambda=$ [-0.660, -0.462, -0.264, -0.198, -0.132, -0.066, 0, 0.066, 0.132, 0.198, 0.264, 0.363, 0.495, 0.693, 0.891, 1.254] \AA. A total of eight frames with an exposure time of 17 ms each were aquired at all wavelength positions in all polarization states. The spectral resolution of SST/CRISP at 5876 \AA\;equals $R=110\,000$. The FWHM of the transmission profile at 5876~\AA\;is $\Delta\lambda=0.052$ \AA. We have sampled the line core of the \hed with $0.066$ \AA\;intervals so we did not critically sample the line, which would need intervals of 0.026 \AA. 

Between 11:57 and 12:01, a total of 15 scans were taken with a cadence of $15$ s, covering the peak time of the flare. The observations were taken in order to test the \hed pre-filter at the SST for the first time, explaining why the observing sequence contains only one spectral line and why the time span is only 3 min 45 s. For the inversions, we have focused on three time steps referred to as $t_{\rm pre}$, $t_{\rm peak}$ and $t_{\rm post}$ (see Fig.~\ref{fig:goes}). We do not have any pre-flare or post-flare observations, $t_{\rm pre}$ and $t_{\rm post}$ refer to pre-peak and post-peak of the flare.

The data have been reduced with the CRISPRED pipeline described in \citetads{2015A&A...573A..40D}. The flatfield images are used without correction for the cavity error, due to the absence of \hed signal in the quiet sun where the flats are taken. The other reduction steps were standard such as the application of image restoration with Multi-Object Multi-Frame Blind Deconvolution (MOMFBD, \citealtads{2005SoPh..228..191V}).

\begin{figure*}
\center
\includegraphics[scale=1]{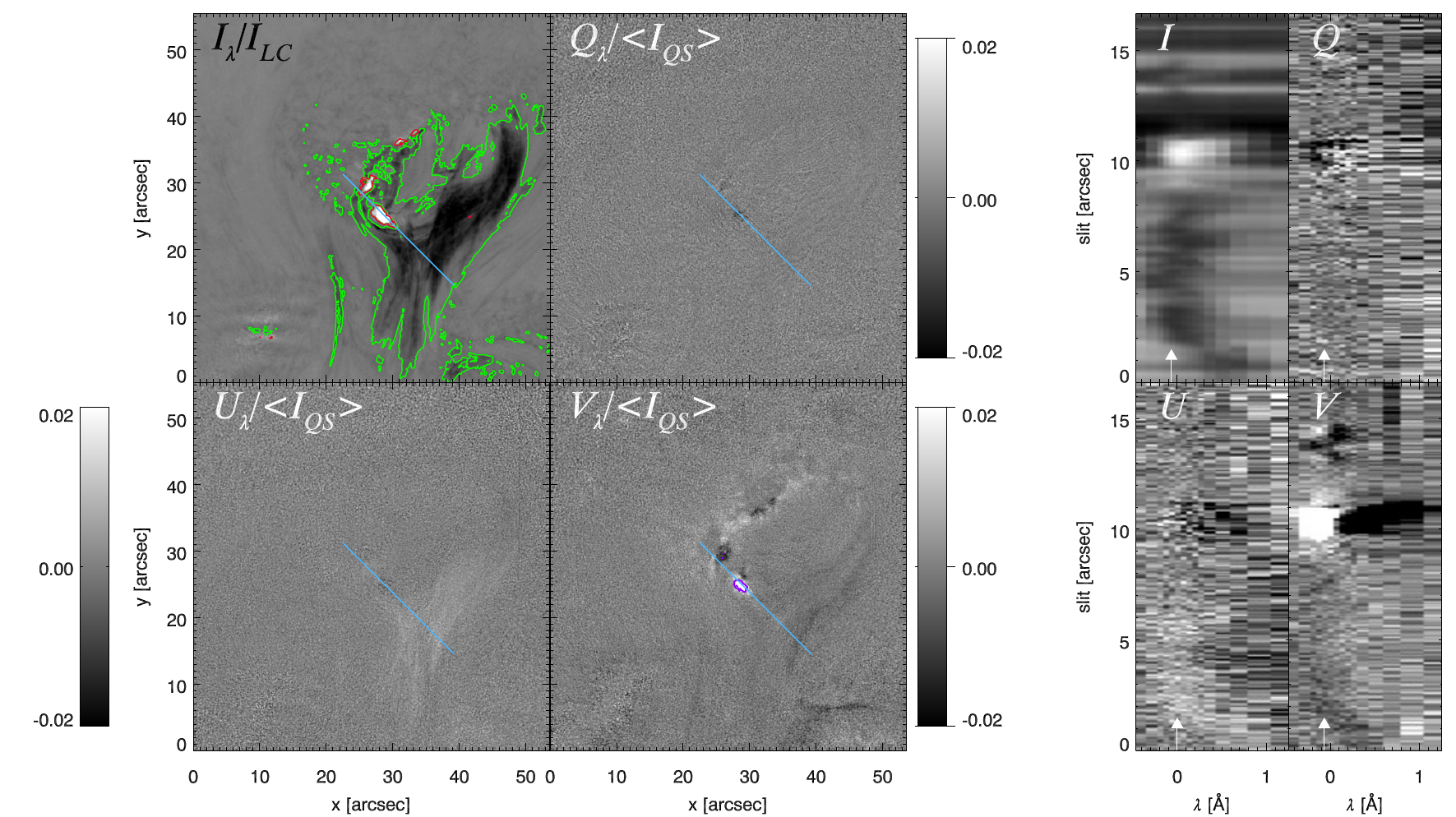}
\caption{Stokes parameters of the in the \hed line at $t_{\rm peak}$. The four panels on the left show the flare images, while the four panels on the right show the flare spectra. {The Stokes $I$ image has been corrected with the local continuum $I_{LC}$. The Stokes $Q$, $U,$ and $V$ images have been divided by an average patch of quiet sun intensity $<I_{QS}>$}. The spectra are extracted along the blue line overplotted on the images. The white arrow on the spectra indicates the wavelength position at which the images are shown. The red contours overplotted on the Stokes $I$ image indicate the areas of \hed emission, the green contours indicate the \hed absorption structure as determined from the k-NN algorithm. {The purple contour overplotted on the Stokes $V$ image indicate areas where the signal is larger than 0.02.}\label{fig:stokes}}
\end{figure*}

Since the polarimetric signal is close to the noise level for many pixels, we corrected carefully for cross-talk, telluric blends and pre-filter imprint. The final data after all corrections is shown in Fig.~\ref{fig:stokes} for $t=~$11:58:23 UT at $\lambda=-0.132\,\AA$. There is a detection of signal in all Stokes parameters in at least some parts of the flare. The noise level is estimated at $~4\cdot 10^{-3}$, this is the standard deviation of a subfield in a quiet sun region in which we assumed the polarimteric signal to be zero. This assumption is likely valid since we obtain similar values at all wavelength positions.


\subsection{Context data of SDO}
Data from the the Atmospheric Imaging Assembly (AIA, \citealtads{2012SoPh..275...17L}) and the Helioseismic and Magnetic Imager (HMI, \citealtads{Schou2012}) on board of the Solar Dynamics Observatory (SDO) were used for context. We have aligned the nearest-neighbor-in-time images of SDO with each SST scan, using some of the routines of Rob Rutten. The original pixel scale of SDO has been preserved. 

Additionally, the HMI full magnetic field vector is available for AR 11235 at the time of our observation, with 720 s cadence. We have used the radial component of the full magnetic field vector projected onto cylindrical equal area (CEA) coordinates for a comparison with the output of \textsc{Hazel}, shown in Fig.~\ref{fig:magcomp}. In all other figures, the line-of-sight HMI magnetogram is used. The difference between both is small, since the heliocentric angle of our observations is $\mu=0.93$. Unfortunately, this event has not been observed by RHESSI, IRIS, or Hinode.

\section{Data analysis}\label{sec:ana}
\subsection{k-NN clustering \label{sec:knn}}

We have applied a k-nearest neighbor (k-NN) algorithm to classify the profiles into three clusters: the background \hed signal which does not correspond to the flaring region, an emission cluster and an absorption cluster. The k-NN clustering algorithm has been successfully used by \citetads{2017A&A...602A..80D} to automatically detect superpenumbral microjets in large datasets and we refer the reader to this paper for a detailed description and background on the k-NN algorithm.

Only one loop over the dataset is required to cluster every profile. The result is shown as red and green contours in Fig.~\ref{fig:stokes}. 83.4\% of the profiles are classified as background, 16.2\% as absorption and 0.4\% as emission. An estimated 0.05\% of the profiles in the scan show both emission and absorption components but we have chosen not to classify those seperately. Therefore, those profiles get classified according to the cluster that contains the most similar seed profiles. 

Generally speaking, the k-NN clustering is successful at distinguishing between background and flare profiles, with some exceptions for for example, a fibril at $x=23\arcsec, y=10\arcsec$ and a small sunspot at $x=40\arcsec, y=2\arcsec$. The background \hed profiles have a mean intensity of $0.95 I_{LC}$ in the line core and no detectable polarimetric signal, with $I_{LC}$ being the local continuum. 


\subsection{Inversions with HAZEL}
\subsubsection{Signal-to-noise ratio and inversion strategy \label{par:sigtonoi}}

In order to infer the magnetic field configuration and thermodynamic parameters of the flare, we fit the \hed profiles using the \textsc{Hazel} inversion code \citepads{2008ApJ...683..542A}. The largest challenge to overcome is the fitting of low signal-to-noise ratio Stokes profiles. In Appendix \ref{app1}, we discuss our attempts to increase the signal-to-noise ratio of the data. Eventually, we opted to apply a spatial rebinning of $2\times 2$ pixels.

We chose to invert the pixels of the absorption and the emission cluster separately, using a different number of free parameters, see Table~\ref{tab:pars}. This decision originated in the fact that there are qualitative differences between the Stokes parameters of the clusters. For the absorption profiles, the Stokes $Q$, $U,$ and $V$ profiles are all of similar order of magnitude, {in other words between $0.1-2\%$ of the mean quiet sun  intensity $<I_{QS}>$}. Since those signals are of the same order of magnitude as the photon noise, we have to be mindful of degeneracy of the fits. Therefore in our opinion, it is wise to fit those profiles with only one slab hence minimizing the amount of free parameters. There is evidence of a second redshifted component in many absorption profiles, with a small imprint on intensity and substantial polarimetric signal. Since we chose not to fit the second component, we have excluded those wavelength points from the inversion by assigning them a lower weight, see the grayshaded area in Fig.~\ref{fig:absprofs}.

{The purple contour in Fig.~\ref{fig:stokes} indicates the area where the emission profiles have strong Stokes $V$ signals with $\frac{V_{\lambda}}{<I_{QS}>}$ up to 15\%. }The strong Stokes $V$ profiles also show a strong peak asymmetry and the corresponding Stokes $I$ emission profiles show evidence of two velocity components. The Stokes $V$ asymmetry could not be fitted properly with a one slab inversion or via inclusion of atomic orientation (see Appendix~\ref{par:ator}). Therefore, the profiles of the emission cluster are fitted with two slabs, each with a different value for the magnetic field. The slabs are put on top of one another and the outgoing radiation of the lower slab is used as a lower boundary condition for the upper slab, see Eq.~4 in \citetads{2017A&A...598A..33L}.
We note that we took the spectral point spread function (PFS) of SST/CRISP into account in the inversions.

\begin{table*}
\begin{center}

  \begin{threeparttable}
    \caption{Fitted parameters of the inversion for the emission and absorption cluster.\label{tab:pars}}
      \begin{tabular}{l l l l}
       
        Cluster & N\textsubscript{pars} & Fitted parameters & Constants \\
        \hline
                Emission & 14 & $B_1$, $\theta_{B,1}$, $\chi_{B,1}$, $\tau_1$, $v_{\rm Dop,1}$, $v_{\rm mac,1}$, $\beta_1$ & $h=2'',~a=10^{-4}$\\
                 & & $B_2$, $\theta_{B,2}$, $\chi_{B,2}$, $\tau_2$, $v_{\rm Dop,2}$, $v_{\rm mac,2}$, $\beta_2$ & \\\hline
        Absorption & 7 & $B$, $\theta_{B}$, $\chi_{B}^{a}$, $\tau$, $v_{\rm Dop}$, $v_{\rm mac}$, $\beta$ & $h=10'',~a=10^{-4}$\\

     \end{tabular}
    \begin{tablenotes}
      \small
      \item  {\bf Notes}. $^{(a)}$ The magnetic field azimuth has been inverted only with one Levenberg-Marquardt cycle, not applying the DIRECT algorithm. The input values for the Levenberg-Marquardt were calculated via loop orientation measurement, see Sect.~\ref{sec:loop} .
    \end{tablenotes}
  \end{threeparttable}
  \end{center}
\end{table*}

\subsubsection{Fitting the absorption profiles: azimuth measurement via fibril detection \label{sec:loop}}
The fact that the Stokes $Q$ and $U$ signal for the absorption profiles is of the same order of magnitude as the noise level introduces difficulties to retrieving a reliable estimate for the magnetic field azimuth. Moreover, up to eight potential ambiguities are present for the azimuth measured via \hed \citepads{2005ApJ...622.1265C,2008ApJ...683..542A}. {Therefore, we made the assumption that the magnetic field is roughly aligned with the thin strand of plasma observed as \hed absorption in the flare loop system. The validity of this assumption has been studied in the context of chromospheric fibrils nearby sunspots via Ca \textsc{ii} 8542 \AA\;\citepads{2011A&A...527L...8D,2017A&A...599A.133A} and via \hei \AA\;\citepads{2013ApJ...768..111S,2015SoPh..290.1607S}. All of the authors have found an agreement between the magnetic field alignment and fibril orientation, though an average error of 10\degree\;is reported by \citeads{2013ApJ...768..111S} while \citeads{2017A&A...599A.133A} found a standard deviation of up to 34\degree\;in weakly magnetized regions. The agreement between the orientation of flare loops and magnetic field has not been studied via spectro-polarimetry so far, but we assume them to be rougly aligned and proceed by giving \textsc{Hazel} an estimate for the azimuth via fibril detection to use as input for the Levenberg-Marquardt algorithm.} This strategy tackles the challenge of noise and the challenge of ambiguities simultaneously and results in clean maps for the magnetic field parameters. In Appendix \ref{app3}, we discuss how the assumption of the magnetic field being aligned with the fibrils affects our results.

\begin{figure*}
\begin{center}
\includegraphics[scale=1]{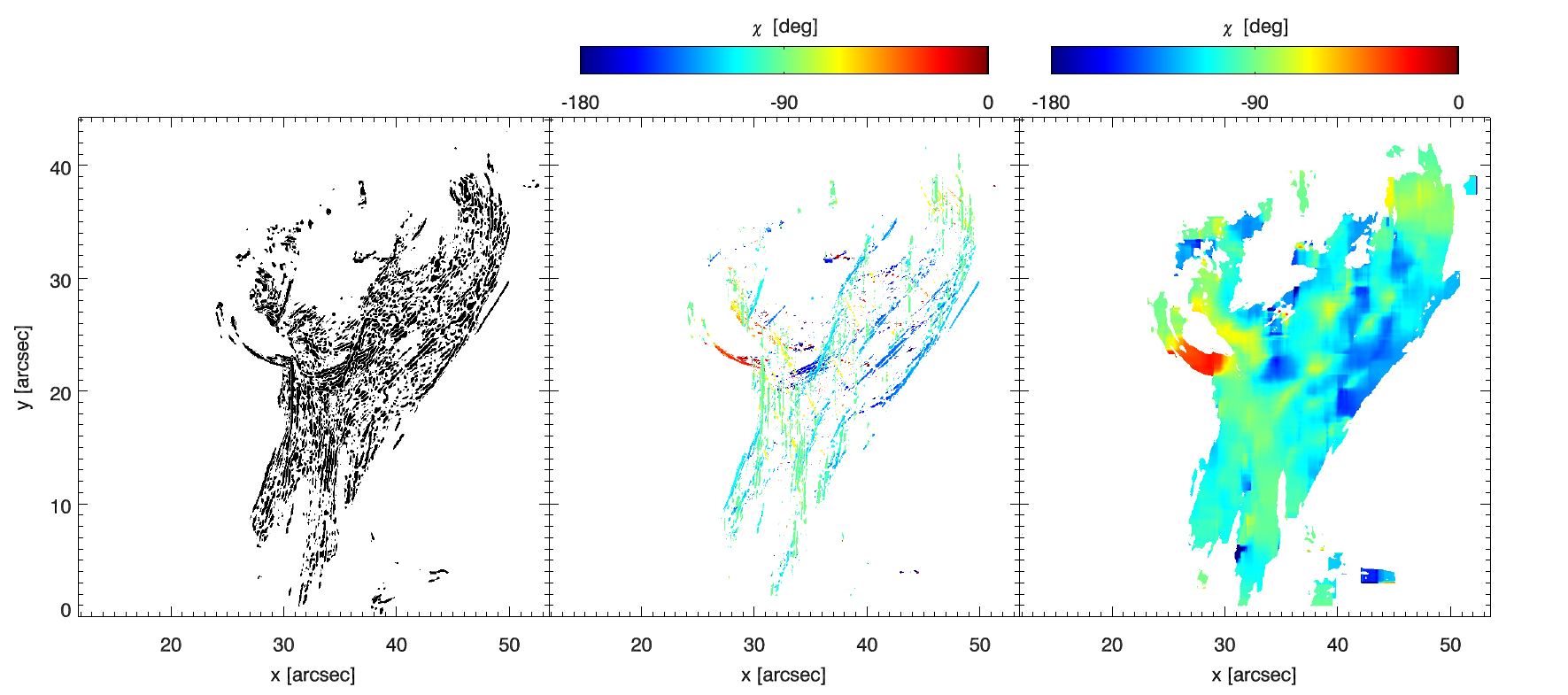}
\caption{Demonstration of the Rolling Hough Transform method to select loops and measure their orientation. Left panel: masked image highlighting linear structure. Middle panel: the selected loops exhibiting a dominant orientation with a measurement of their orientation. Right panel: a smoothed version of the middle panel, used as an input for the azimuth in our inversions. \label{fig:loops}}
\end{center}
\end{figure*}

We detected the flare fibrils and measured their orientation using a method called rolling Hough transform as was presented by \citetads{2014ApJ...789...82C} and applied to chromospheric fibrils by \citetads{2017A&A...599A.133A}. We followed the method closely as described in Sect.~3 of \citetads{2017A&A...599A.133A} and apply it to our non-binned continuum corrected data. After some testing, we use parameter values $D_K=5$ which is the width of the top-hat smoothing kernel, $D_W=30$ which is the radius of the subfield used to select fibrils, and $Z=1.8 \cdot \sqrt{ \frac{N_a}{\pi D_w^2}}$ being the cutoff value for selection of fibrils (see \citealtads{2017A&A...599A.133A} for a detailed explanation of the meaning of these parameters). $N_a$ equals the number of pixels corresponding to a linear structure selected by the bitmask (see left panel of Fig.~\ref{fig:loops}) and within a circle of radius $D_W$. We made $Z$ variable since a method with a fixed value for $Z$ would prefer to select fibrils that were surrounded by many other fibrils within a radius $D_W$, while fibrils at the edge of our structure were disadvantaged by the selection method.

The result of the rolling Hough transform is shown in Fig.~\ref{fig:loops}. The left panel shows the application of a bitmask that is obtained by substracting a smoothed image from the original one, highlighting linear structure in the image. The middle panel shows the selected fibrils and the measurement of their orientation. The figure in the rightmost panel shows a smoothed version of the fibril orientation that we use as an input for our inversions. We have only used this input value for the azimuth to fit the absorption profiles, not the emission profiles. Only the Levenberg-Marquardt algorithm was used to fit the azimuth, starting from the loop orientation measurement as an input. For all other parameters of the inversion we have used a combination of the DIRECT algorithm  and Levenberg-Marquardt \citepads{2008ApJ...683..542A}.

\subsubsection{Quality of the fits}
\begin{figure*}
\begin{center}
\includegraphics[scale=1]{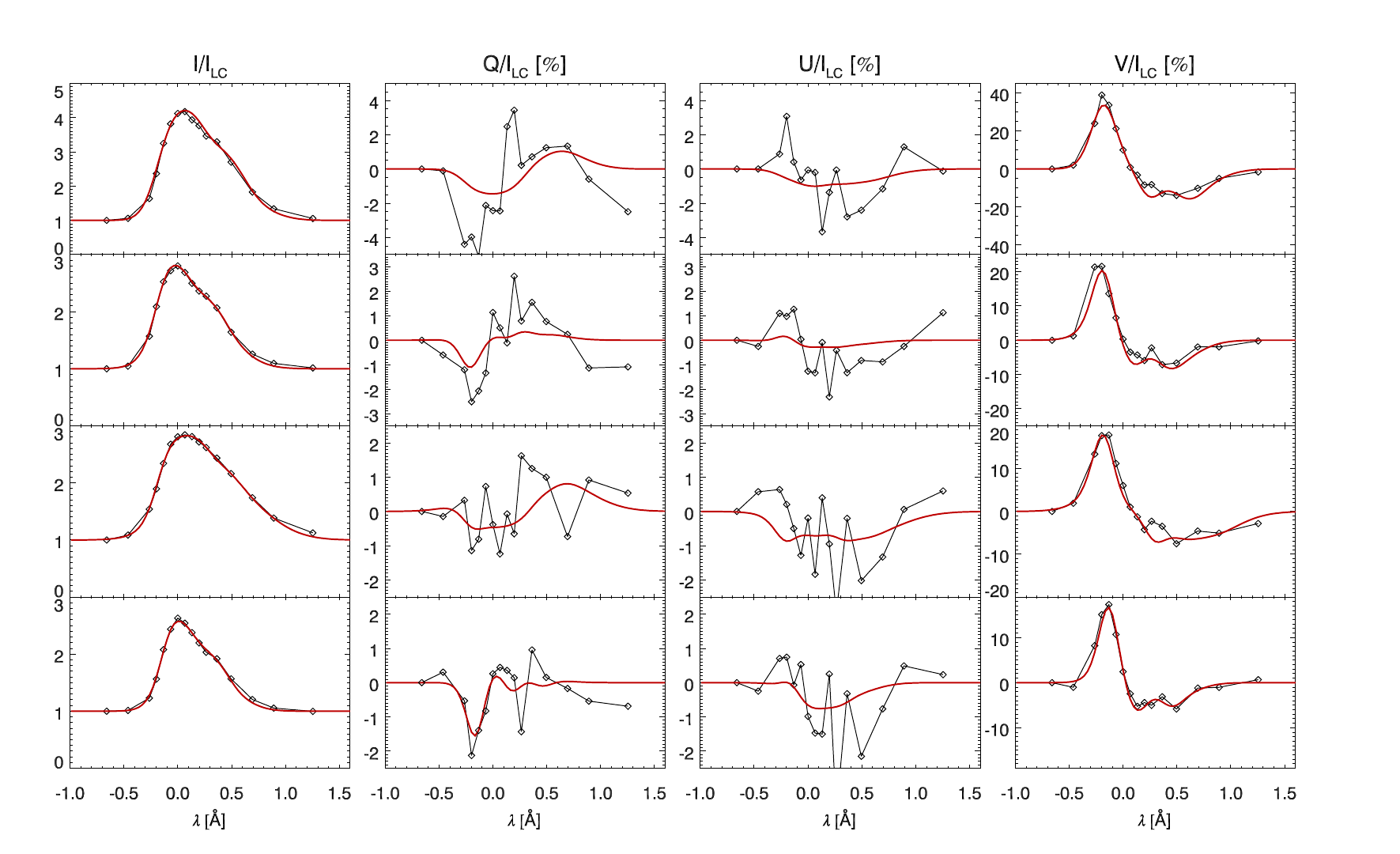}
\caption{Some examples of the quality of the two-component fits to the Stokes parameters of the \hed emission profiles. Black diamonds indicate the observed wavelength points, the black line connects the observed points and the red line indicates the fitted result. {All profiles are normalized using the local continuum $I_{LC}$.}\label{fig:emprofs}}
\end{center}
\end{figure*}

\paragraph{Emission profiles} {Figure \ref{fig:emprofs} shows some examples of emission profiles, normalized with the local continuum $I_{LC}$. We note that those strong Stokes $V$ signals of up to 40\% are in part due to the fact that they are located above a dark pore so that the value of the local continuum $I_{LC}$ is small. As a comparison, the polarization percentages for the top profile in Fig.~\ref{fig:emprofs} equal
max($Q_{\lambda}/I_{\lambda},U_{\lambda}/I_{\lambda}, V_{\lambda}/I_{\lambda}) = (0.9\%, 2.7\%, 16.2\%)$.} The quality of the two-component fits to Stokes $I$ and Stokes $V$ of the emission profiles is satisfactory (without the introduction of atomic orientation, see Appendix \ref{par:ator}). Therefore, we trust that we have inferred the line-of-sight velocity $v_{\rm LOS,1}$ and $ v_{\rm LOS,2}$ of the two slabs with a reasonable accuracy. Since Stokes $V$ is generated by the longitudinal Zeeman effect, we trust that we have at least a good order of magnitude estimation for the magnetic field strength $B_1$ and $B_2$ and the magnetic field inclination $\theta_{B,1}$ and $\theta_{B,2}$. The Stokes $Q$ and $U$ signals are likely generated within the saturated Hanle regime, meaning that Stokes $Q$ and $U$ are only sensitive to magnetic field orientation and not to its strength. The fits to Stokes $Q$ and $U$ are not satisfactory at the footpoint location and therefore, we do not trust that we have determined the azimuth of the magnetic field correctly. The reason is a combination of things: the Stokes $Q$ and $U$ profiles might have very complex shapes due to multiple components (see e.g.,~\citealtads{2011A&A...526A..42S} where the flaring profile show evidence of up to five components). This complexity in combination with the noise level and the fact that we have not critically sampled the profiles leads to profiles that are hard to interpret and fit. 

We did not use an input value for the azimuth as determined from loop orientation for the emission profiles. Nevertheless, we still have an azimuth that has a similar value to its surroundings: without imposing any starting value, the azimuth in the emission profiles at $x=28'',y=25''$ converged to a value of around $\simeq -45$ deg, similar as the small loop southward of the emission profiles. It might be that \textsc{Hazel} still converges to a realistic solution in a statistical way, even though the individual profiles are too noisy to deduce information from them.

\begin{figure*}
\begin{center}
\includegraphics[scale=1]{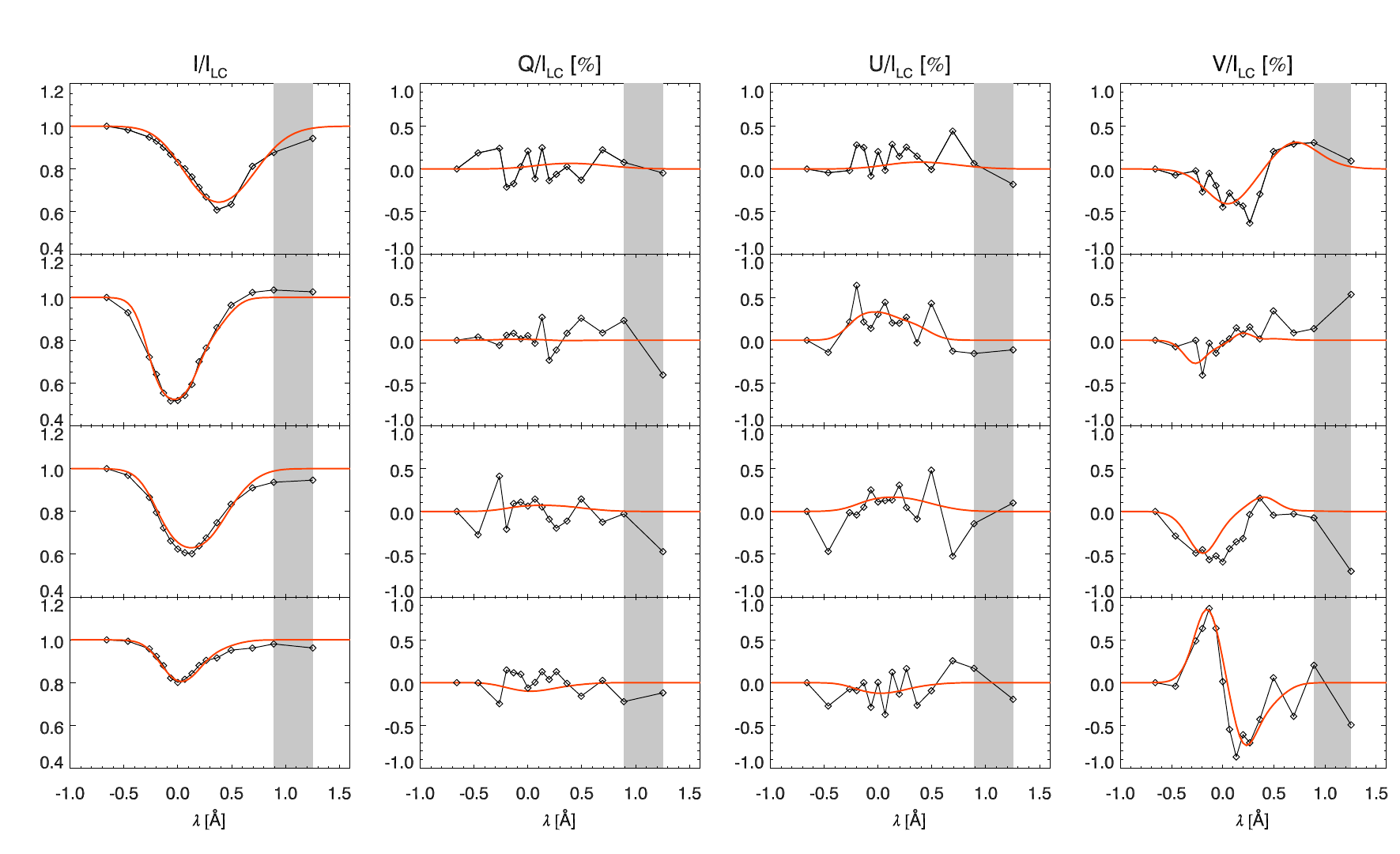}
\caption{Some examples of the quality of the fits to the Stokes parameters of the \hed absorption profiles. Black diamonds indicate the observed wavelength points, the black line connects the observed points and the red line indicates the fitted result. The gray shaded area marks the area that is excluded from the inversions. {All profiles are normalized using the local continuum $I_{LC}$.}\label{fig:absprofs}}
\end{center}
\end{figure*}

\paragraph{Absorption profiles} The quality of the one-component fits to the absorption profiles is demonstrated in Fig.~\ref{fig:absprofs} where some examples of fits are shown. The pixels were selected for their diversity of the profiles, not because of the quality of the fit. Those four examples can hence be taken as typical average quality fits, representative for the rest of the inverted profiles. As mentioned in Sect.~\ref{par:sigtonoi}, we have chosen to exclude wavelength point at $\Delta\lambda=$ [0.891, 1.254] \AA\;from the fit. Apart from the red wing of the absorption profile, the core of Stokes $I$ is fitted well using \textsc{Hazel}, meaning that we have determined the line-of-sight velocity $v_{\rm LOS}$ of the line core accurately. The Stokes $V$ signal is generated by the longitudinal Zeeman effect, but the quality of the fits is hard to judge due to the high noise level. For Stokes $V$, we claim that we fit the polarity of the field correctly and that we obtain an order of magnitude estimate for the magnetic field strength. 

The azimuth as obtained from the inversion is very similar to the input calculated from the fibril orientation. When trying random values as input for the azimuth, we find that \textsc{Hazel} always converges to a solution that is close to the input value (see Appendix \ref{app3}). Therefore, it is clear that the observations do not provide sufficient information to determine the azimuth reliably, and we cannot use the value of the azimuth in our interpretation of our results. However, as is demonstrated in Fig.~\ref{fig:absprofs}, \textsc{Hazel} has made an attempt to fit the $Q$ and $U$ profiles. Due to the fact that \textsc{Hazel} performs inversions in the local vertical reference frame and not in the line-of-sight frame, the value of the azimuth does affect the value for the inclination and magnetic field. We cannot disentangle the line-of-sight magnetic field without knowing the azimuth. In Appendix \ref{app3}, we describe the study we performed of the sensitivity of the magnetic field strength and inlination to the azimuth and we find that the value of the azimuth has limited effect on {the magnetic field strength and non-negligible affect on the inclination. We obtain qualitatively similar results for the magnetic field strength in all cases. The inclination is affected by our choice of azimuth, however the polarity of the field is still the same in many areas of the flare, though not in all.} Our assumption of the magnetic field being aligned with the fibrils does hence not overly influence the results - regardless of whether the assumption is valid or not.

\subsubsection{Coordinate system of \textsc{Hazel}}
The coordinate system of \textsc{Hazel} is set up with respect to the local vertical, which is the natural coordinate frame for using slab geometry and calculating the Hanle effect. The angles $\theta_B$ and $\chi_B$ that we obtain from the inversions are the angles between the magnetic field vector and the local vertical (See Fig.~1 of \citealtads{2008ApJ...683..542A}). $\theta$ is the heliocentric angle between the local vertical and the line of sight. 




\section{Results}\label{sec:res}

\subsection{Description of \hed signal and line formation \label{sec:lineform}}

\paragraph{Emission profiles \label{par:emission}}

The brightest \hed features are the emission kernels in the flare, see the red contours in Fig.~\ref{fig:stokes}. The two largest emission kernels are located at coordinates $x=28''$,$y=25''$ and at $x=27''$,$y=30''$. Both kernels have a diagonal of around $\simeq 3''$ along their longest axis. 

\begin{figure*}
\includegraphics[scale=1]{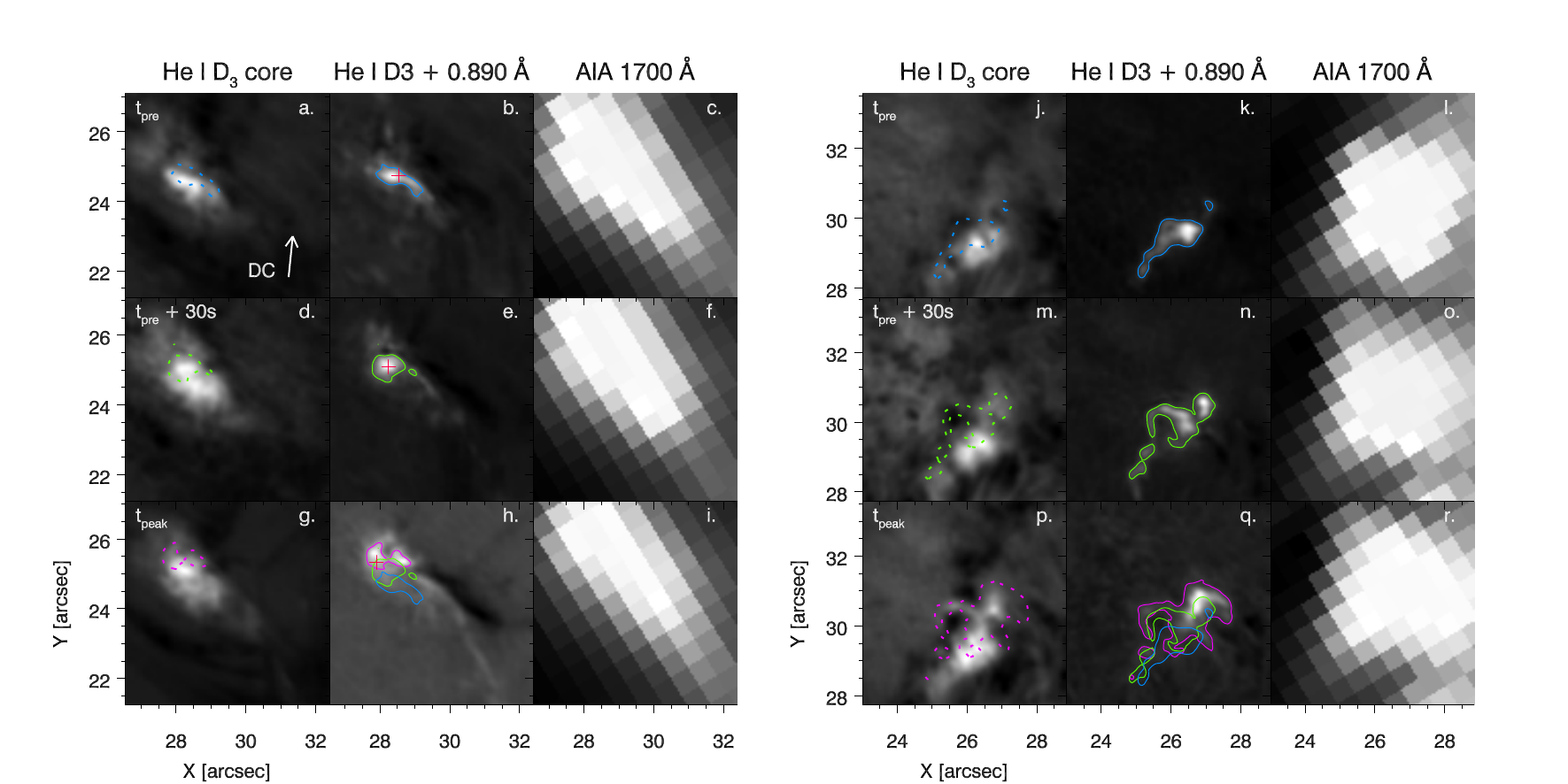}
\caption{Magnification of the two largest \hed emission kernels in the data at three different time steps $t_{\rm pre}$,$t_{\rm pre}+30s$ and  $t_{\rm peak}$ . An image in the \hed line core an at a redshifted wavelength point of $\Delta \lambda=890\;\rm m\AA$ is shown. Co-aligned SDO 1700 \AA~images are shown as well, with the original pixel scale preserved. The red plus marks in panels b, e, and h correspond to the profiles displayed in Fig.~\ref{fig:ta}. The solid line contours in the images at the \hed red wing show an intensity threshold in that image. The same contours are overplotted as dashed lines in the \hed line core images and in panels h and q to demonstrate a time and space lag.  \label{fig:emission}}
\end{figure*}

\begin{figure*}
\includegraphics[scale=1]{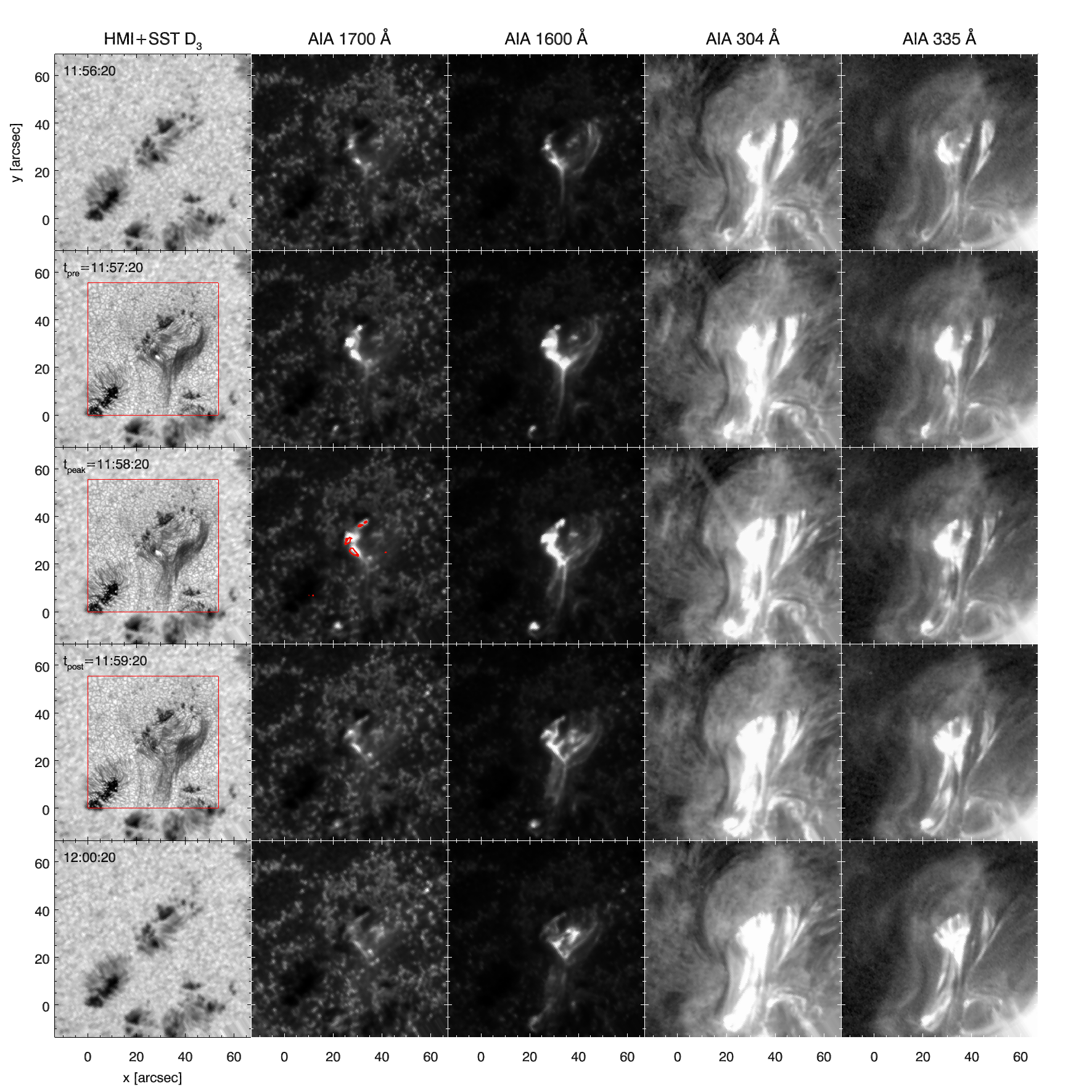}
\caption{Context for the C3.6 class flare from SDO/AIA and SDO/HMI. From left to right we display SDO/HMI continuum, SDO/AIA 1700 \AA, SDO/AIA 1600 \AA, SDO/AIA 304 \AA\,and SDO/AIA 335\AA. From top to bottom we display five different time steps as indicated in the left corner of the images in the first column. In the first column, we have overplotted the SST field-of-view at the \hed line core $\lambda=0$ \AA, at the time steps where it was available, as indicated by the red box. We indicated the \hed emission contours corresponding to the SDO/AIA 1700 \AA\;emission areas. \label{fig:bigfov}}
\end{figure*}

A magnification of the two emission kernels is shown in Fig.~\ref{fig:emission} for three different time steps. Figure~\ref{fig:emission} demonstrates that the emission kernels move with apparent plane-of-the-sky velocities of 10--20$\, \rm km\,s^{-1}$. The two kernels are located at opposite sides of the polarity inversion line (PIL). The largest emission kernel shown in panels a--i of Fig.~\ref{fig:emission} moves in a parallel direction along the PIL while the emission kernel shown in panels j--r moves perpendicularly away from the PIL. Co-spatially with \hed emission, we observe strong brightenings in the SDO 1700 \AA\;channel of up to a factor seven times the mean intensity of the SDO 1700 \AA\;channel in a quiet area in the FOV. The \hed emission kernels coincide one-to-one with the strong brightnenings in the SDO 1700 \AA\;channel, demonstrated by the red contours in Fig.~\ref{fig:bigfov}. 

The emission kernels are locations of chromospheric condensations, which we discuss in Sect.~\ref{sec:chromcond}. We interpret the \hed emission kernels as the locations of the flare footpoints, defined by non-thermal electrons emitting Bremsstrahlung in hard X-rays (HXR). Those non-thermal electrons might be populating the \hed levels via direct collisional excitation and/or ionization-recombination cascades. Thermal processes might be relevant too. We observe strong broadening in the emission profiles which is a combination of many effects: velocity gradients, macroscopic turbulence and a part of the broadening must also be thermal. However, it is hard to pinpoint how large the thermal contribution is. \textsc{Hazel} does not calculate temperature and it is hard to model temperatures for \hed because there is always the unknown contribution from photoionizing (E)UV photons, and in this case non-thermal electrons.

Unfortunately RHESSI was not observing during the time of our observation, so we cannot directly link the locations of HXRs with the location of \hed emission. However, we provide the following evidence for the \hed emission kernels being related to the flare footpoints. Firstly, the \hed emission kernels coincide one-to-one with SDO/AIA 1700 \AA\;channel brightenings (see red contours in Fig.~\ref{fig:bigfov}). The 1700 \AA\;channel is dominated by the Balmer continuum for temperatures larger than 10\,000 K \citepads{2016A&A...590A.124R}. Brightenings of the Balmer continuum are found to coincide with flare footpoints \citepads{2014ApJ...794L..23H,2017ApJ...836...12K,2017ApJ...837..125K,2017ApJ...837..160K}. However, a recent paper by \citetads{2018arXiv180801488S} has suggested that the 1700 \AA\,channel has only minor contribution from the continuum. Secondly, \hed emission has been linked with the presence of HXRs \citepads{1981ApJ...248L..45Z,1983ApJ...271..832F,2013ApJ...774...60L}. Lastly, velocities of the apparent plane-of-the-sky motion of the emission kernels agrees with estimates of footpoint motions of flares in the literature, see for example,~\citetads{2005AdSpR..35.1707K} and \citetads{2007ApJ...654..665T}.

\paragraph{Absorption profiles}

The most striking feature of the \hed flare observation is the dark absorption structure seemingly floating over the photospheric granulation. The absorption is spatially sub-structured with fine elongated dark fibrils of plasma. The fibrils are overlapping at certain locations increasing the \hed opacity, suggesting an optically thin regime. The absorption structure stays present througout our observation, also after the impulsive phase of the flare, but is fading a little over time. No large-scale dynamics is obvious from the \hed images.

The \hed absorption profiles are often broad and show a variety of velocity shifts from modestly blueshifted to strongly redshifed, see Sect.~\ref{sec:vel}. Some of the profiles show evidence of two velocity components along the LOS. 

All EUV channels of SDO/AIA show bright emission co-spatially to the \hed absorption structure, see Fig.~\ref{fig:bigfov}. All channels are strongly overexposed except the AIA 335 \AA\;channel and the AIA 304 \AA\;channel which we chose to display. The SDO/AIA EUV channels also reveal a dark band at $x=40'',y=10$--$45''$ in the 304 \AA\;and the 335 \AA\;channel displayed in Fig.~\ref{fig:bigfov}. The dark band might represent cool material being pushed upwards by a heated chromosphere.

The shape of the \hed absorption structure and the SDO/AIA emission structure is comparable to a ``$Y$'' or upside-down Eiffel tower. This type of events are often observed on smaller scales where they are called anemone jets, and are associated with reconnection events. The reconnection likely takes place in the location where the legs of the structure meet: at around ~ $x=32'',y=25''$ in Fig.~\ref{fig:bigfov}, most clearly visible as a bright blob in the 1600 \AA\;channel. In the time series of SDO/AIA images, the large scale dynamics of the flare becomes obvious. The top of the Eiffel tower connects outside the SST field-of-view at location $x=21'',y=-7''$\footnote{We note that what we call the top of the Eiffel tower or the top footpoint is in fact located in the bottom of the figure, southward of the SST FOV.}. This footpoint manifests itself as a bright kernel in all SDO/AIA UV and EUV channels. The 335 \AA\;and the 304 \AA\;channel show bright blobs of plasma traveling from the reconnection site to this footpoint.

The geometry of the event suggests that the flaring structure rises high in the solar atmosphere. We can derive this from projection properties. We have assumed that the top and bottom footpoints are formed at comparable height in the solar atmosphere, probably in the deep chromosphere. The flaring structure has a size of $\sim 40\arcsec$ in the plane of the sky. At an average  inclination of, for example, 45 deg, the apex of the structure would reach $\sim 14$ Mm up in the atmosphere, comparable to prominence heights.

Our interpretation of the \hed absorption is that they trace the flare loops. The loops are located high in the atmosphere and connect to the flare footpoints, formed at lower altitudes exhibiting \hed emission. The \hed absorption is either created via collisions caused by thermal conduction or by photoinization-recombination. The emission in SDO/AIA channels suggest that a large part of the flaring plasma is at coronal temperatures. However, sufficient neutral helium is still present to provide \hed opacity. There are two possibilities: the first possibility is that a strong non-equilibrium regime is present in which the helium ionization temperature and the electron temperature are decoupled \citepads{2014ApJ...784...30G}. Considering that the target is flaring renders a strong non-equilibrium regime quite plausible. Another possibility is that the coronal emission is not emitted at the same altitude in the solar atmosphere as the \hed absorption. Possibly, the \hed absorption is originated in a region right below the hottest loops that are emitting in the EUV at the location of older loops. Photoionization-recombination would then be a likely mechanism to populate the \hed levels.

\subsection{Chromospheric condensations \label{sec:chromcond}} 
Figure \ref{fig:emission} compares the emission in the \hed line core and in the red wing of the \hed line, revealing an interesting behavior. The emission kernels have a ``preceding'' narrow front ($\sim 0.3''$) of redshifted profiles in the direction of motion, followed by the appearence of a larger emission area radiating in the \hed line core. These narrow fronts are indicated in Fig.~\ref{fig:emission} as contours in the \hed wing images and are overplotted in the \hed line core images to show that they are preceding the line core radiation in both time and space. This process has been described a few times before in the literature: chromospheric downflows observed in hydrogen Balmer lines, Ca \textsc{ii} K or Mg \textsc{ii} h\&k occur at the leading edge of flare kernels \citepads{1980IAUS...91..217S,1997A&A...328..371F,2015ApJ...807L..22G}. We demonstrate this effect for the first time in the \hed line. The SST/CRISP data offers \hed spectra with high temporal cadence (15 s) and full spatial coverage of the emission kernels, necessary to study the moving and quickly evolving \hed emission kernels.

\begin{figure*}
\includegraphics[scale=1]{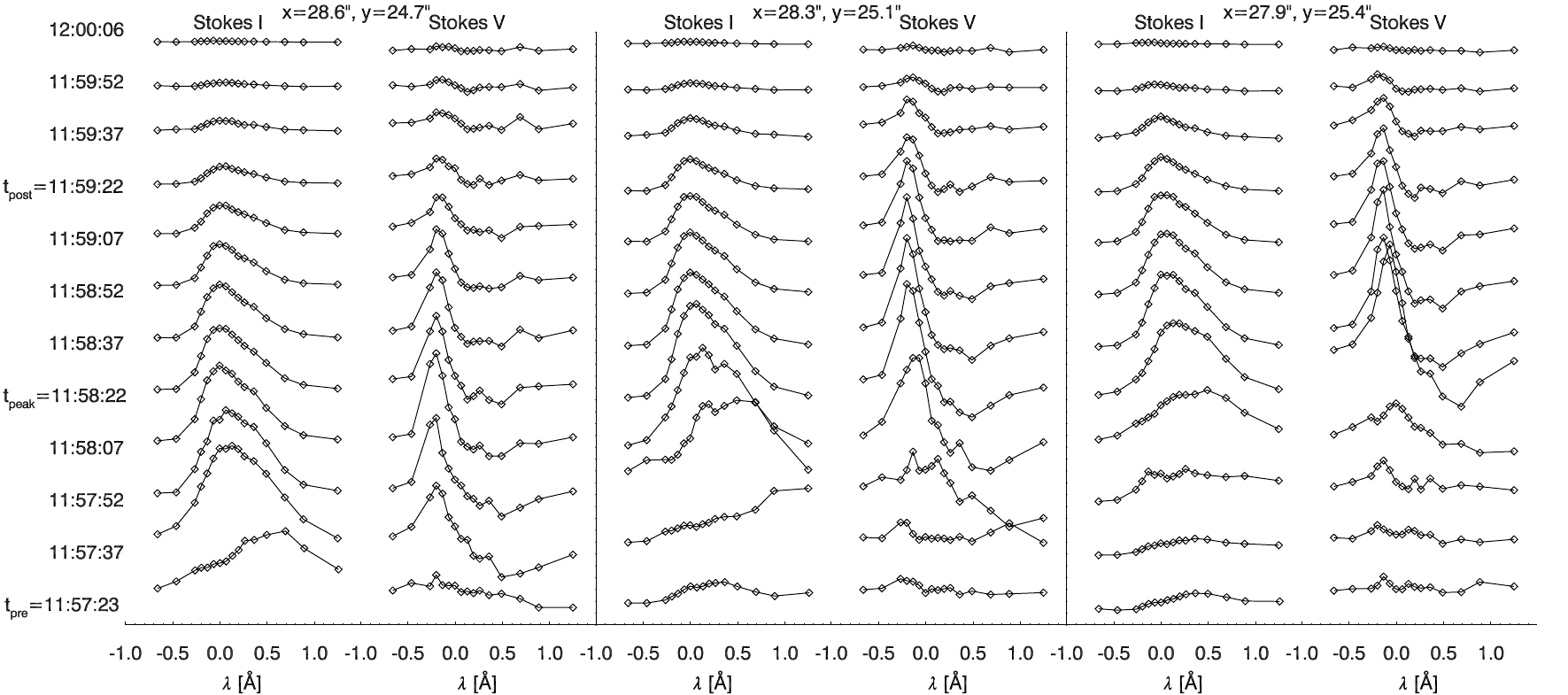}
\caption{Time evolution of the Stokes $I$ and Stokes $V$ parameter of the \hed profile in three pixels, indicated by red plus signs in Fig.~\ref{fig:emission}. The diamonds correspond to the wavelength points measured in our observation. \label{fig:ta}}
\end{figure*}

Our observation is suggestive of plasma flows with high downwards velocity impacting the stationary atmosphere below and causing a shock. Figure~\ref{fig:ta} illustrates the time evolution of this process for three different pixels. All profiles show similar temporal evolution, even though the process is not acting co-temporaly in those pixels. At first, the \hed profile is predominantly redshifted, sampling the downward plasma flow. In the following time steps, the emission increases, broadens and moves toward more stationary to even blueshifted profiles that are sampling a shocked atmosphere. Simultaneously, the Stokes $V$ signal becomes very strong as soon as the stationary velocity component is present suggesting formation height in the deep chromosphere for the \hed line. At this height, the magnetic field is expected to be strong since the flare occurs above photospheric pores and penumbral area. Two simultaneous velocity components are obvious in the Stokes $I$ profiles at multiple time steps. The emission is strongest during the impulsive phase of the flare and has decreased substantially by $t_{\rm post}$. In conclusion, we observe both strong chromospheric condensations (downflows) and modest upflows in the \hed line. The observed condensations are preceding the upflows - we suspect that the modest upflows are caused by the chromospheric condensations, see a discussion in Sect.~\ref{sec:disccond}. 

\begin{figure}
\includegraphics[scale=1]{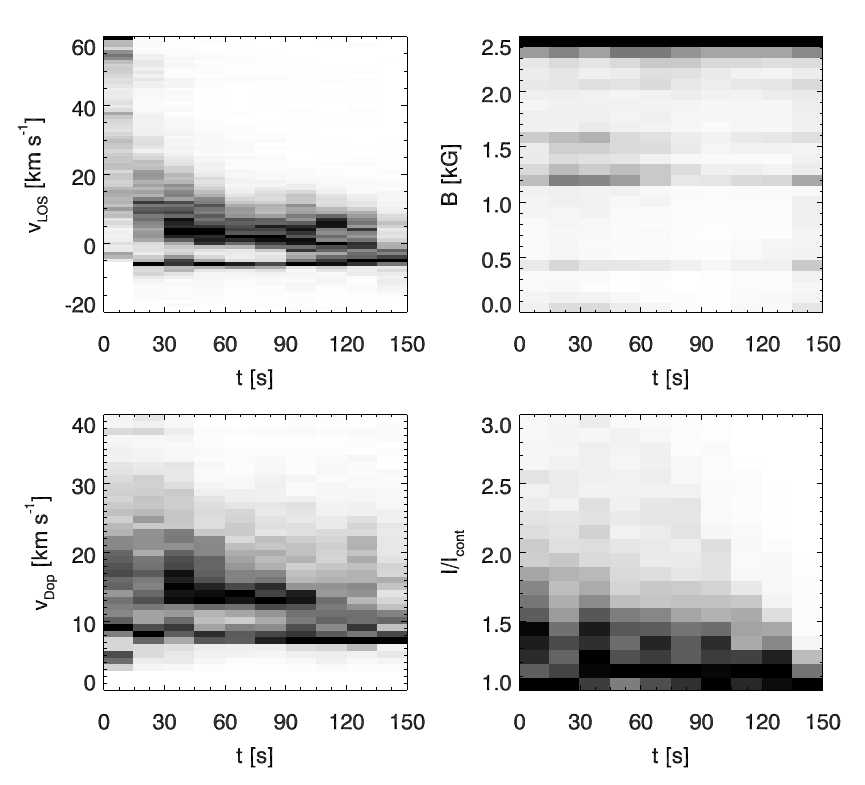}
\caption{Time evolution of the chromospheric condensations, measured by \textsc{Hazel}. The time bin is 15s, corresponding to the observed cadence. Each time bin is normalized to its maximum value. We show the line-of-sight velocity $v_{\rm LOS}$ (top left), the magnetic field strength B (top right), the Doppler broadening $v_{\rm Dop}$ (bottom left) and the continuum corrected intensity of the profile $I/I_{\rm cont}$ (bottom right).  \label{fig:evap}}
\end{figure}

We have analyzed the chromospheric condensations in all available flare time steps using \textsc{Hazel} and we performed a mean epoch analysis inspired by \citetads{2015ApJ...807L..22G}: the profile with the highest redshifted velocity was put at $t=0$ s. The subsequent time steps are then put into the subsequent time-bins of 15 s, since that corresponds to the cadence of our observation. The result is shown in Fig.~\ref{fig:evap} for the LOS velocity $v_{\rm LOS}$, the magnetic field $B$, the Doppler width $v_{\rm Dop}$ and the intensity as compared to the continuum intensity $\frac{I}{I_{LC}}.$ 

The decay time of the chromospheric condensations is around $\rm t\simeq 60$ s during which the line-of-sight velocity diminishes from $\sim 60\,\rm km~s^{-1}$ to $\sim 5 \rm\,km\, \rm s^{-1}$. The presence of the second blueshifted component is striking in this diagram as well: on average its LOS velocity is around $\sim -5\,\rm km~s^{-1}$. The blueshifted component becomes much more pronounced at $\rm t=15$ s, one time step after the largest chromospheric condensations are measured, suggesting that the chromospheric condensations cause the observed upflows.

No time-evolution of the magnetic field is obvious from Fig.~\ref{fig:evap}. The observed variation of Stokes $V$ is caused by the intensity evolution of Stokes $I$ and not by the evolution of the magnetic field. We measure very strong values of up to $B\sim 2500$ G at the locations of the \hed emission. {We note that we have set the upper limit of the inversion range for $B$ to 2500 G, see Sect.~\ref{subs:mag} for a discussion)}. The somewhat discrete appearance of the value for $B$ is caused by the use of the DIRECT fitting algorithm in \textsc{Hazel} (see \citealtads{2008ApJ...683..542A} for a description of the DIRECT algorithm).

The Doppler broadening shows a similar but noisier time evolution to the line-of-sight velocity and is also suggestive of two components: one broad component with $v_{\rm Dop}\simeq 15\;\rm km~s^{-1}$ and one component with $v_{\rm Dop}\simeq 8\;\rm km~s^{-1}$. However, one should not interpret the Doppler broadening parameter too literal. Since we performed constant slab inversions with \textsc{Hazel}, the Doppler broadening parameter is used by the code to attempt to fit any unmodeled effects: velocity gradients, multiple line components, combination of emission and absorption profiles and any type of broadening that is not necessarily related to Doppler motions. In our case, the broadened appearance of the redshifted component is likely caused by a combination of velocity gradients, macroscopic turbulence and thermal Doppler motions.

\subsection{Maps of thermodynamic parameters\label{sec:vel}}

\begin{figure}
\center
\includegraphics[trim={3.5cm 0 0 0},clip,scale=1]{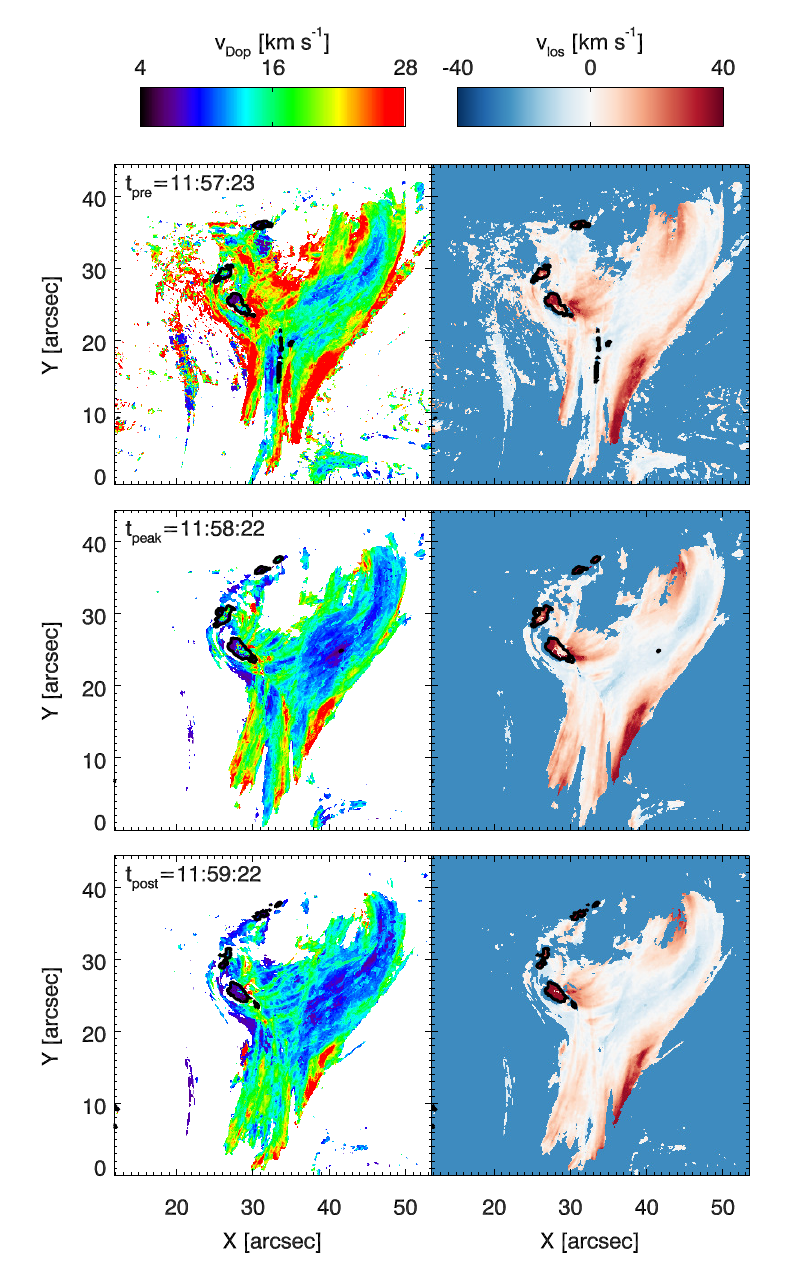}
\caption{Inversion results on $2 \times 2$ rebinned data of the thermodynamic parameters: the Doppler broadening $v_{\rm Dopp}$ (left column) and the line-of-sight velocity $v_{\rm LOS}$ (right column). The three rows represent three different time steps: $t_{\rm pre}$, $t_{\rm peak}$ and $t_{\rm post}$. The black contours indicate the locations of the \hed emission profiles. The emission profiles have been inverted using two components, of which the redshifted component is shown here. \label{fig:therm}}
\end{figure}

Figure \ref{fig:therm} shows the inversion results of the thermodynamic parameters the Doppler broadening $v_{\rm Dop}$ and the line-of-sight velocity $v_{\rm LOS}$. At $t_{\rm pre}$, the Doppler broadening is clearly a lot larger in many locations in the flare than in $t_{\rm peak}$ and $t_{\rm post}$. At $t_{\rm pre}$, more profiles are selected by the kNN-algorithm to be part of the flare. This is in small part due to worse seeing at $t_{\rm pre}$, but mostly due to intrinsic properties of the plasma. More pixels in the FOV show substantial \hed opacity at $t_{\rm pre}$ and the profiles are generally speaking much more complex at $t_{\rm pre}$ in comparison to $t_{\rm peak}$ and $t_{\rm post}$. There is evidence multiple velocity components and velocity gradients. As mentioned in Sect.~\ref{par:emission}, these effects are fitted in \textsc{Hazel} by increasing the Doppler broadening parameter.

Since locations of high Doppler broadening are indicating locations of ``unmodeled'' physics, it is clear that we are not capturing all information present in the data, especially at $t_{\rm pre}$. We had already suggested that the absorption profiles often show evidence of a redshifted velocity component along the line-of-sight, mostly present at the early times of the flare. However, the spectral sampling and spectral coverage of the data in combination with the signal-to-noise ratio of the data, does not allow us to unambiguously include a second component in the inversions without the introduction of to many free parameters and hence strong degeneracy of the results. We also observe that high values of Doppler broadening seem to correlate with high values for the LOS velocity, which is not surprising if we interpret high Doppler broadening as likely locations of velocity gradients. 

The maps for the line-of-sight velocity in Fig.~\ref{fig:therm} reveal a complex velocity field for the flare structure. Several locations exhibit strongly redshifted profiles corresponding to downflows along the LOS on the order of $\sim~30$--$40~\rm km~s^{-1}$. A strong downflow at $x=30'',y=25''$ is located right next to the largest and brightest emission kernel (shown in panel a--i of Fig.~\ref{fig:emission}). This connects to our interpration of the emission kernels as the footpoints of the flare, exhibiting large downflows. Paragraph~\ref{par:emission} also discussed those downflows causing shocks in the lower atmosphere at the location of the emission kernels. The measured velocity map is consistent with those findings. 

We also measure strong downflows at $x=28'',y=12''$ and $x=35'',y=15''$ in the ``top'' part of the Eiffel tower structure where loops connect outside the SST field of view. We note that at the same location, the hot SDO/AIA channels exhibit a plasma flow from the reconnection site toward the top footpoint at $x=21'',y=-7''$. We cannot measure the line-of-sight velocity for SDO/AIA but likely, those observed hot plasma flows would yield negative line-of-sight velocities, hence upflows toward the observer. Observing simultaneous upflows in hot lines and downflows in cool lines is a signature of explosive chromospheric evaporation.

The bulk of the loop structure is stationary to weakly blueshifted with velocities between $-15$ and  $0~\rm km~s^{-1}$. The measured velocities are strongest in the first two time steps and are decreasing toward $t_{\rm post}$, as is the intensity of the observed profiles. 


\subsection{Maps of the magnetic field}\label{subs:mag}

\begin{figure*}
\center
\includegraphics[scale=1]{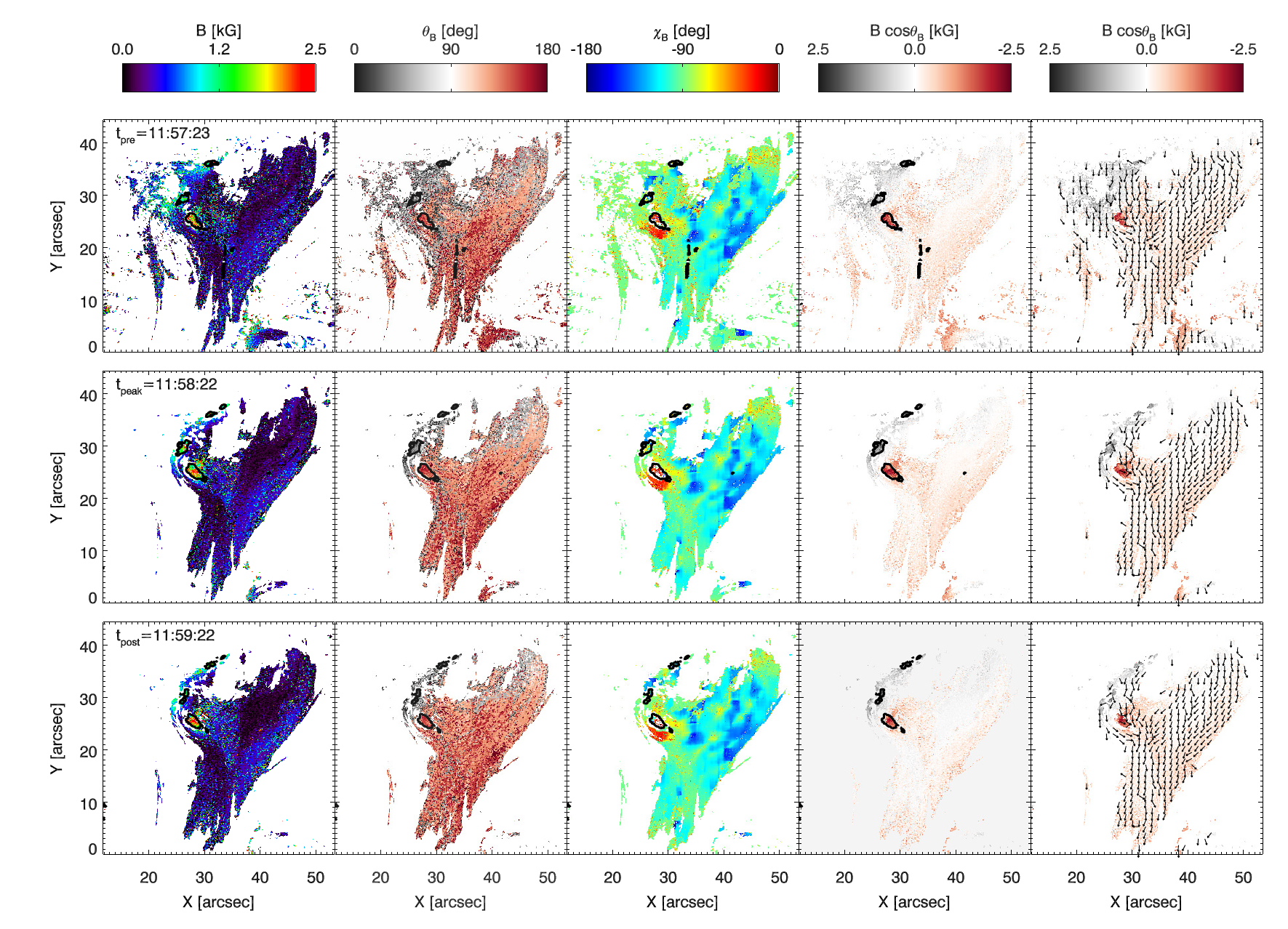}
\caption{Inversion results on $2 \times 2 $ rebinned data of the magnetic field with respectively from left to right: the magnetic field value $B$, the inclination $\theta_B$, the azimuth $\chi_B$ (input value determined by fibril orientation), the vertical magnetic field $B\cdot \cos\theta_B$ and the vertical magnetic field with arrows overplotting the azimuth direction. The three rows represent three different time steps, top to bottom: $t_{\rm pre}$, $t_{\rm peak}$ and $t_{\rm post}$. The black contours indicate the emission profiles. The emission profiles have been inverted using two components, of which the blue component is shown here.\label{fig:inv} }
\end{figure*}

\begin{figure}
\center
\includegraphics[scale=1]{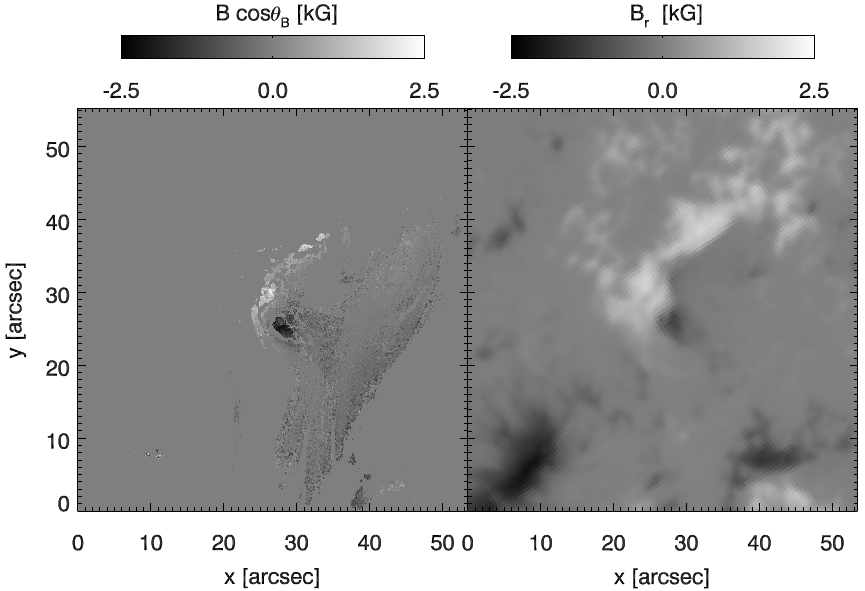}
\caption{Comparison of the vertical magnetic field as obtained from \textsc{Hazel} and the radial (=vertical) component of the magnetic field projected onto cylindrical equal area (CEA) coordinates, downloaded as such from the HMI database. \label{fig:magcomp}}
\end{figure}

\paragraph{Emission profiles}
As discussed in detail in Sect.~\ref{par:ator}, the \hed emission profiles exhibit strong and asymmetric Stokes $V$ profiles. Those can be fitted using two components along the line of sight with different $v_{\rm LOS}$ and $B$. The magnetic field strength in the emission kernels of \hed and surroundings has large values for $B$ of up to 2500 G, see Fig.~\ref{fig:inv}. In fact, when increasing the fitting boundaries of \textsc{Hazel} to for example,  $5000$ G, magnetic field strenghts between $2500$--$5000$ G are found as well. However, we applied Occam's razor and set the boundary to the lowest value required to fit the strong Stokes $V$ profiles. Considering that the formation height of the emission kernels is probably located in the low chromosphere to even photosphere, and considering that we are located above pores and a forming penumbral area, high values for the magnetic field are perhaps to be expected. The measured values are substantially higher than in SDO/HMI (see Fig.~\ref{fig:magcomp}) but that could be due to the major difference in spatial resolution and the differences between the spatial PSF of SST and SDO. 

We compared the magnetic field in the flare footpoints between the stationary to modestly blueshifted velocity component and redshifted velocity component of the emission profiles (we refer to them as blue and red components onwards). We find that the blue component has on average a more vertical magnetic field. The inclination angle of the blue component peaks around $\theta_B\sim 30\degree$ and $\theta_B\sim -150\degree$. The red component has its largest peak of the inclination angle distribution at $\theta_B\sim 90\degree$ and two smaller peaks at $\theta_B\sim 30\degree$ and $\theta_B\sim -150\degree$. We suspect that \textsc{Hazel} inverted the red component with a predominantly horizontal magnetic field because we have excluded two red wing wavelength points from the fit. There seems to be evidence of polarimetric signal, also for the red component, but we do not have the spectral sampling in order to be able to fit this signal. This might also explain why the blue component has a magnetic field that is on average 430 G stronger than the red component, at least at times $t_{\rm pre}$ and $t_{\rm peak}$. An additional reason that the blue component has a stronger and more vertical magnetic field is that this component might be effectively formed at lower height in the atmosphere than the red component, where stronger and more vertical magnetic fields are expected. This is consistent with our interpretation of a downflow impacting the deep atmosphere, see Sect.~\ref{par:emission}, and with our interpretation of the emission kernels as flare footpoints. The polarity of the field agrees with the polarity given by the line-of-sight magnetogram of SDO/HMI, see Fig.~\ref{fig:magcomp}. The magnetic field polarity shows concentrated negative polarity at $x=27'',y=26''$ which is surrounded by positive polarity. 

\paragraph{Absorption profiles}

The absorption profiles provide the main part of the maps shown in Figure \ref{fig:inv}. The bulk of the flare has a low magnetic field strength, varying between $0$--$100$ G. This is an expected result because it is located above granulation. There is some large scale variation displaying higher values for the magnetic field of up to $\sim 500$ G in the region for example at $x=40'',y=20''$. This region correlates spatially with the region of high redshifted velocities of up to $v_{\rm LOS}\sim 40 \,\rm km~s^{-1}$ and a more vertical magnetic field  with $\theta_B\sim 170\degree$.

The magnetic field reaches high values of more than $\sim 1000$ G in areas close to the flare footpoints. These \hed absorption profiles are probably generated at lower heights in the solar atmosphere than the bulk of the flare absorption profiles. 

There are no obvious differences in the measurement of the magnetic field strength and inclination between the three inverted time steps, at least in the flare loops. However, close to the flare footpoint at $x=28'',y=27''$, some changes are apparent in the magnetic field between $t_{\rm pre},t_{\rm peak}$ and $t_{\rm post}$. This is consistent with the findings of \citetads{2017ApJ...834...26K} who found that the chromospheric field mostly changes near the flare footpoints. 


\section{Discussion \label{sec:disc}}

\subsection{\hed signal and line formation}
\hed imaging data has been available from BBSO since the seventies, hence it has been known since then that flares have a response in \hed \citepads{1980ApJ...235..618Z}. However, a common idea states that the \hed line goes into emission only in large flares \citepads{1988assu.book.....Z}. This is a misconception, probably due to the limited spatial resolution available at the time the idea was first published. Our observations show that both \hed emission and absorption can be present - also in small flares. Whether there is absorption or emission signal depends on the physical conditions governing the local plasma and the line formation mechanism, not on the strength of the flare.

There are a few studies that have studied helium line formation in flares, although mostly the He \textsc{i} 10830 \AA~line. \citetads{2005A&A...432..699D} have studied the effects of an electron beam on the formation of He \textsc{i} 10830 \AA. They suggest that He~\textsc{i} 10830 \AA\;is strongly enhanced in the non-thermal case and that strong absorption at the flare onset followed by strong emission during the flare maximum would be a signature of non-thermal effects in flare footpoints. We do indeed observe strong emission in the footpoints during the flare maximum but we have not observed absorption in the flare footpoints at an early flare phase. However, our observation started after the flare onset, so we might have missed possible early \hed absorption. The results of \citetads{2005A&A...432..699D} point to a formation mechanism in which non-thermal electrons collide with neutral helium and cause population in the triplet system via ionization-recombination. According to \citetads{2005A&A...432..699D}, this mechanism can lead to strong \hei \AA\; at the flare maximum. We believe this could be a valid mechanism for the observed \hed emission in the flare footpoints.

\citetads{2014ApJ...793...87Z} study He \textsc{i} 10830 \AA\;emission kernels in imaging data from BBSO, connected to the footpoints of a C-class flare that resemble our \hed emission signals. \citetads{2014ApJ...793...87Z} conclude that the photoionization-recombination is most likely to explain their observations of He \textsc{i} 10830 \AA~emission, since their estimated photon budget is consistent with their observations. We however think that the PRM is not so likely to be responsible for the generation of the \hed emission in our observation, since that emission is so localized, contains fine spatial structure, and evolves so quickly in time. On the other hand, we do not know whether there might be sources of EUV photons available in the transition region that evolve in the same morphological way - possibly generating \hed population via PRM (as explained in more detail by \citealtads{2016A&A...594A.104L}). 

The PRM might however very well be responsible for the larger absorption structure that we observe in our dataset. \citetads{2015ApJ...809..104A} include synthesized He~\textsc{i} 10830 \AA~profiles in their models containing XEUV backwarming and without. They do obtain He~\textsc{i} 10830 \AA~absorption of around 25\% of the continuum intensity in their model including XEUV backwarming, as opposed to only $~\sim 5\%$ in their model without including XEUV backwarming. Their calculated absorption level is of similar magnitude as our observed \hed absorption structure. In the model of \citetads{2015ApJ...809..104A}, the levels in the helium triplet system are populated via the PRM mechanism.

We argue that more detailed modeling of helium line formation in flares is necessary to exclude or confirm any line formation mechanisms. In addition, co-observation with IRIS, and for example SST/CHROMIS might provide more clues on the line formation mechanism. It would be interesting to investigate if there is any local heating present, if there are there signs of non-thermal effects and if they are sufficient UV and EUV photons present in the transition region possibly generating \hed population via PRM.

\subsection{Chromospheric condensations \label{sec:disccond}}
The chromospheric condensations measured via \hed, shown in Fig.~\ref{fig:evap} have at the same time confirmed the results of \citetads{2015ApJ...807L..22G} and added extra information to them. The temporal evolution and the decay time of the chromospheric condensations is remarkably similar when measured via the subordinate Mg \textsc{ii} triplet at $2798.7$ \AA\;or via He \textsc{i} D\textsubscript{3}: the velocities decay from $\sim 60\,\rm km~ s^{-1}$ to $\sim 5\,\rm km~s^{-1}$ in $50$--$60$ s. However, something that we observe in \hed and has not been reported by \citetads{2015ApJ...807L..22G} is the presence of a second stationary to blueshifted component appearing $\sim 15$ s after the largest downflows are measured (see Figs.~\ref{fig:ta} and \ref{fig:evap}). We have interpreted this component as sampling a shocked layer in the deep atmosphere impacted by the initial downflow. The reason why we might observe this in \hed and not in the Mg \textsc{ii} triplet is perhaps because the Mg \textsc{ii} triplet is not sensitive to these particular plasma conditions: Mg \textsc{ii} might be ionized away or the core of the line might simply not be sensitive to this formation height. 

\begin{figure}[b]
\begin{center}
\includegraphics[trim={0 2.8cm 0 2.5cm},clip,scale=0.27]{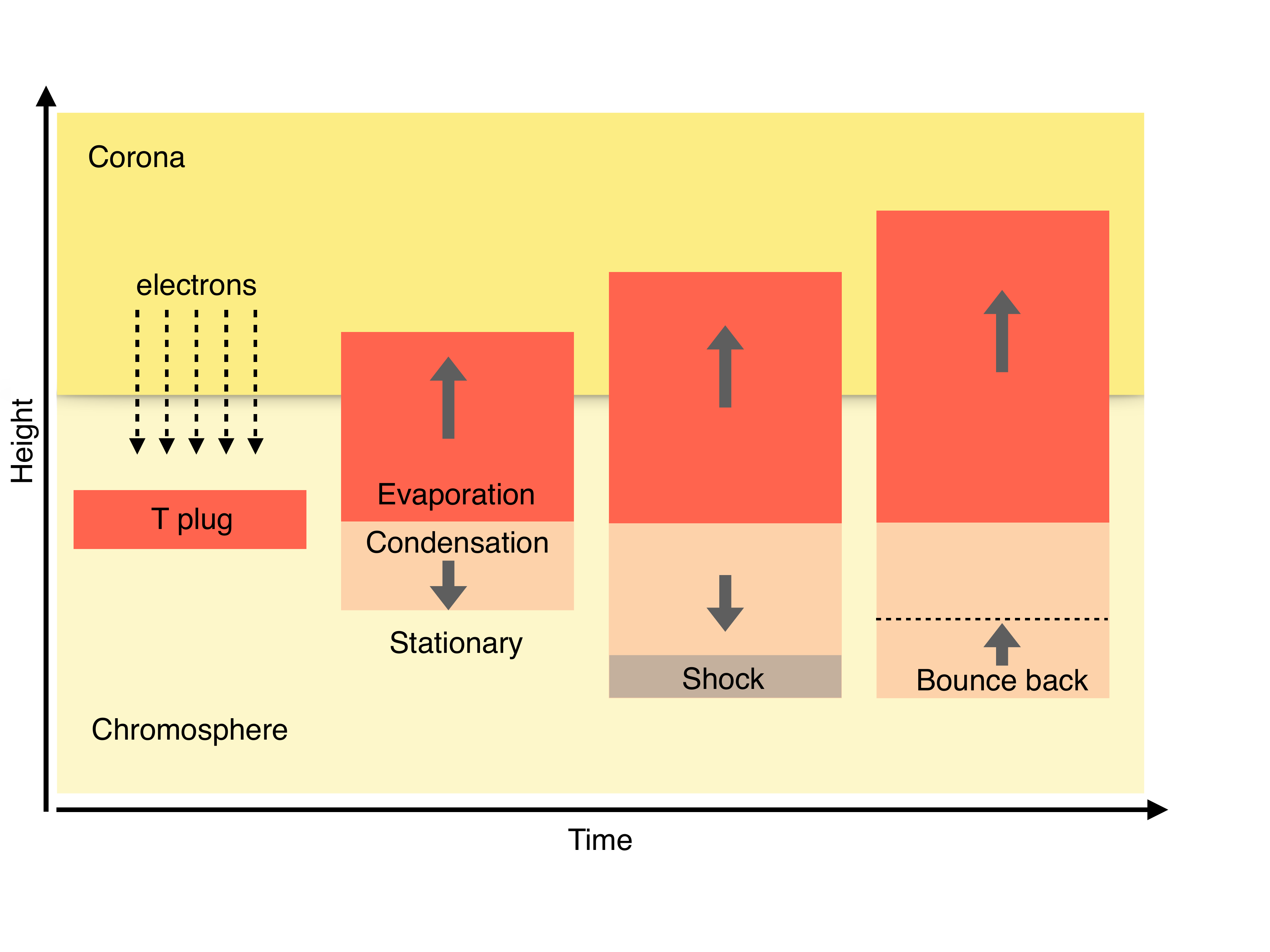}
\caption{Schematic view of a scenario that could possibly explain our observed \hed profiles during chromospheric condensations at flare footpoints. See Sect.~\ref{sec:disccond} for a detailed description. \label{cartoon_cc}}
\end{center}
\end{figure}

Several studies that observe the Mg~\textsc{ii} h\&k lines mention blue asymmetries and/or blue velocities that are present before the largest redshifts are observed \citepads{2015A&A...582A..50K,2015ApJ...807L..22G,2018PASJ..tmp...61T}. If these blue asymmetries can be interpreted as upflows, that result is more or less the opposite of what we observe in \hed. In our observation, the upflows appear after the largest redshifts are measured. This might suggest the presence of more than one layer with upflowing velocities at different temperatures. Co-observations of \hed and Mg~\textsc{ii} h\&k would certainly shed more light on this apparent inconsistency. \citetads{2018PASJ..tmp...61T} have proposed a scenario in which the upflows in Mg~\textsc{ii} h\&k are caused by a cool upflow on top of the evaporated layer (see Fig.~12 in \citealtads{2018PASJ..tmp...61T}). We have not found any evidence of such upflow in the \hed line, but that could be because the \hed is not sensitive to the cool temperatures ($10^4$ K) of said layer.

Figure~\ref{cartoon_cc} proposes a qualitative scenario to give a possible explanation for our \hed observations. The first step of this scenario displays a classic electron bombardment with non-thermal electrons causing a hot temperature plug in the mid-chromosphere. We do not comment here on the electron acceleration mechanism or the heating mechanism since detailed modeling should decide which mechanism is compatible with our \hed observations. The following time step shows typical chromospheric evaporations and condensations. Chromospheric evaporation corresponds to upflows of up to $-200\,\rm km~s^{-1}$ which are observed in hot lines (e.g., Fe \textsc{xxi} by \citealtads{2015ApJ...807L..22G}). Chromospheric condensation is observed as downflows on the order of $50\,\rm km~s^{-1}$ in cool chromospheric lines and in the redshifted component of \hed. The third time step displays our proposed mechanism in which the chromospheric condensations compress the deep chromosphere, causing a shock. Subsequently, this layer might get heated and will ``bounce back'' with a modest velocity of $\sim -5 \rm km~s^{-1}$ when the chromospheric condensations have decreased in velocity. The layer in which the shock is caused could then explain the observed broadened, enhanced and stationary to modestly blueshifted component in our \hed profiles.

A similar, though not exactly equal, scenario has been proposed by \citetads{2017ApJ...836...12K} to explain white light and NUV emission at flare footpoints. \hed emission has been found to bare a spatio-temporal coincidence with white light and Balmer continuum emission in flares \citepads{1980ApJ...235..618Z,1981ApJ...248L..45Z}. Therefore, it might be plausible that \hed and white light in flares are emitted in the same layer in the atmosphere, or caused by a common physical mechanism. Perhaps the scenario proposed in Fig.~\ref{cartoon_cc} could serve as a qualitative attempt to find common physical ground for \hed and white light emission. Further investigation is certainly worthwhile and would be served well by co-observations of \hed and the NUV with for example,~IRIS. 

In any case, the fact that we observe so clearly the impact of the downflows on the deep chromosphere is unique and a strong argument in favor of observing \hed and He~\textsc{i} 10830 \AA\,during flares when one is interested at studying the chromosphere. \hed is formed in a layer where both downflows and modest upflows happen, rendering the line invaluable to test different models for heating transport and electron acceleration. \citetads{2016ApJ...827..101K} have for example modeled the Mg \textsc{ii} k and the Ca \textsc{ii} 8542 \AA~lines in two different models for chromsopheric heating during flares. We suspect the \hed line to be an extremely useful addition to this set of diagnostics.

\subsection{Chromospheric inversions in flares}
The number of studies providing chromospheric velocity and magnetic field maps of flares is limited. \citetads{2014A&A...561A..98S} have inverted an activated filament during a flare using spectro-polarimetric observations of the He~\textsc{i} 10830 \AA~line using \textsc{Helix} \citepads{2004A&A...414.1109L,2007AdSpR..39.1734L}. They obtained magnetic field and velocity maps that are comparable to our results. The velocities range from $-20$--$50 \rm\,km~s^{-1}$ and show similar behavior as in our observation: more redshifted toward the edges of the structure while the bulk of the absorption is moving upwards. The magnetic field map shows values between $B\simeq 0$--$500$ G, also similar to our results. \citetads{2014A&A...561A..98S} however do not measure the high values for the magnetic field of up to $\sim 2500$ G. One of the reasons might be that they excluded emission profiles from their inversions \citepads{2011A&A...526A..42S} or another cause for finding lower magnetic fields could be that their flare is not located above similarly strong pores and penumbral area.

It is more difficult to compare inversion results originating from Ca \textsc{ii} 8542 \AA~observations. The line has a different appearance in flares as compared to \hed so the maps look substantially different as well. The LOS velocity map of \citetads{2017ApJ...846....9K} shows downflows in the flare ribbons of around $\simeq 15\,\rm km~s^{-1}$. However, the rest of the flaring structure also exhibits mostly weak downflows as measured via the  Ca \textsc{ii} 8542 \AA~line. The LOS velocity map shown in \citetads{2018arXiv180500487K} mostly harbours weak upflows in the ribbon area. The longitudinal magnetic field has values of up to $\simeq 1500$ G. The obtained maps via Ca \textsc{ii} 8542 \AA~have a substantially noisier appearance compared to ours. This might be caused my many things: differences in polarimetric signal-to-noise ratio, \hed does not have a self reversal so it is easier to measure velocities, and NICOLE performs node-based inversions while \textsc{Hazel} uses constant slabs.

\citetads{2017ApJ...834...26K} has used the weak field approximation for Ca~\textsc{ii} 8542 \AA\;to estimate chromospheric magnetic field changes in an X-class flare. The chromospheric magnetogram also exhibits some strong magnetic field of up to $\simeq 1500$ G. She mostly measures magnetic field changes in the chromosphere near flare footpoints, in agreement with our observations. The lack of time-evolution in the flaring loops could actually mean that the magnetic field is rather stationary in that area. Other possible reasons are that the change in magnetic field due to reconnection has already taken place by the time this observation started, or the height at which the field changes in not sampled by the \hed line, or our inversion results are not sufficiently precise to catch any magnetic field changes.

The most significant conclusion of this is that no one so far has measured as high values for the magnetic field in flare footpoints as we did. Fig.~\ref{ao2c} demonstrates that the fits to Stokes $V$ in the flare footpoints are acceptable and indicate that high values of the magnetic field are indeed present. In Sect.~\ref{par:ator} we have excluded the possibility that atomic orientation was causing such strong Stokes $V$ profiles. We claim that \hed in the flare footpoints is probably formed in the deep chromosphere or at even lower heights. Also, the high spatial resolution of the observation and hence the limited smearing by the spatial PSF might play a role in finding such high values for the magnetic field \citepads{2013A&A...557A..24V}.

\subsection{Magnetic field toplogy \label{magtop}} 

Analysis and inversions of the data have revealed several aspects of the flare that can be put together to propose a magnetic topology consistent with the following findings:
\begin{enumerate}
\item The \hed emission kernels are corresponding to the flare footpoints, formed in the deep chromosphere or even lower.
\item The \hed absorption is tracing the flare loops, formed higher in the atmosphere.
\item The deepest formed component of the flare footpoints have a strong and inclined magnetic field with $B \sim 2500$ G and $\theta_B\sim30/150\degree$. The redshifted component of the flare footpoints which is likely formed higher in the atmosphere has a slightly weaker and perhaps a more horizontal magnetic field, see Sect.~\ref{subs:mag}.
\item A strong negative polarity is surrounded by positive polarity.
\item The magnetic field in the loops is much lower with values between $B\simeq 0$--$500$ G. Also the field is on average more horizontal than in the footpoints with $\theta_B\sim 50/130\degree$.
\item We observe strong downflows in \hed of up to $\sim 40\,\rm km~s^{-1}$ close to the footpoints of the flare and close to the reconnection site.
\item We have assumed that the magnetic field follows the fibril direction. In Appendix \ref{app3}, we show that the data does not contain the information to test the validity of the assumption. However, we can conclude that the value of the azimuth does not greatly affect the values for the magnetic field strength and inclination.
\end{enumerate}

All the above findings are suggestive of a fan-spine magnetic field configuration for the flare, as been observed in many flares \citepads{2001ApJ...554..451F,2013ApJ...769..112D,2016A&A...591A.141J,2018ApJ...854...64S} and anemone jets \citepads{1994ApJ...431L..51S,2007Sci...318.1591S,2010ApJ...722.1644S,2017ApJ...844...28S,2016ApJ...819L...3Z}. Many (R)MHD simulations have also modeled these type of events \citepads{2009A&A...494..329M,2010A&A...512L...2A,2013PASJ...65...62T,2016ApJ...822...18N}.

\begin{figure}
\center
\includegraphics[scale=0.8]{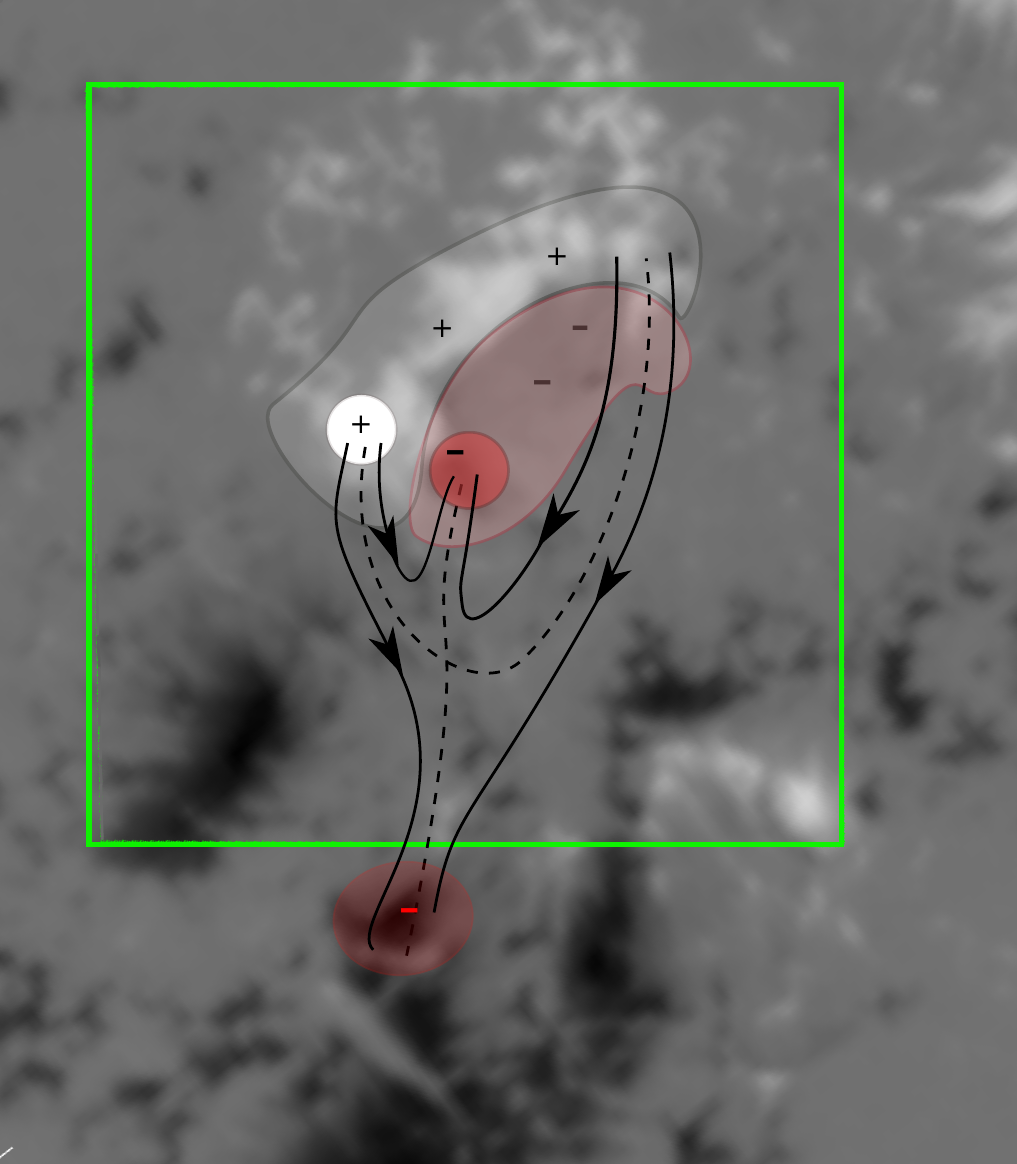}
\caption{Cartoon of the proposed fan-spine magnetic field topology. The background shows the SDO/HMI magnetogram. The green box indicates the SST field-of-view. The red areas show negative polarity kernels, corresponding to the flare footpoints, while the white areas show positive polarity in the flare footpoint area. Suggested magnetic field lines are given in black. The dashed lines indicate the fan-spine configuration. \label{fig:cartoon}}
\end{figure}

In the nominal fan-spine magnetic field configuration, a circular polarity inversion line (PIL) is present and the magnetic field lines are dome shaped. Our observation embodies only a part of this configuration: the PIL describes roughly half of a circle and the field lines also form only a half dome. In Fig.~\ref{fig:cartoon}, we show a simplified cartoon of the proposed fan-spine magnetic field configuration that is consistent with the HMI magnetogram (Fig.~\ref{fig:magcomp}) and with our inversion results (Fig.~\ref{fig:inv}). 

Some similar events are described in the literature, for example \hei imaging observations of a chromospheric jet by \citeads{2016ApJ...819L...3Z}. The authors claim based on imaging data that the magnetic field topology is a fan-spine configuration. The appearance of the event in the \hei line is very similar to the appearance of the \hed line in our observation. A difference is that our observation is almost at disk center while those of \citetads{2016ApJ...819L...3Z} are very close to the limb. Therefore, the structure observed by \citetads{2016ApJ...819L...3Z} is almost entirely observed in emission because of a darker background (keep in mind that both \hed and \hei are limb brightened). Also, the authors suggest that the reconnection of the jet is located in the chromosphere, because many of the hot SDO/AIA channels do not have a large signature of the jet. In our case the flare reconnection likely occurs at coronal temperatures since all SDO/AIA channels have clear imprints of the C3.6-class flare.

\citetads{2013ApJ...774...60L} have also proposed a circular dome fan-spine magnetic field configuration for a flare for which they used \hed imaging data from 1984 observed at BBSO. The magnetic field configuration was retrieved from force-free magnetic field extrapolations.

We have explored the magnetic field in the C3.6 class flare with potential magnetic field extrapolations using the SDO/HMI magnetogram as input. However, the connectivity of the field lines did neither agree with the appearance of the flare in SDO/AIA channels, nor with our inversion results. Non-linear force free magnetic field extrapolations are probably needed for a more realistic result, but this is outside the scope of the current work.

\subsection{Recommendations for flare observations of \hed}
The observations presented in the current work were taken as test observations for a new \hed pre-filter for SST/CRISP, bought in 2015 and replaced with a higher-quality pre-filter in 2016. Therefore, the observational program was not tailored to capture flares. For future flare observations employing \hed with SST/CRISP (or any Fabry-P\'erot instrument at other telescopes), we recommend to choose between aiming for the study of either chromospheric condensations, or chromospheric magnetic field.

For the study of chromospheric condensations, we recommend choosing a program with a very large spectral range to capture extreme velocities, see for example the highly redshifted profiles in the middle column of Fig.~\ref{fig:ta}. We recommend a spectral range including downflows of up to at least $100\,\rm km~s^{-1}$ and to have a dense wavelength sampling of the \hed line region. By sacrificing polarimetry, the cadence can be kept high, which is necessary to capture the quick temporal evolution of the chromsopheric condensations.

For the study of the chromospheric magnetic fields, we recommend to sacrifice temporal cadence and opt for a large spectral range, dense sampling of the spectral region to capture the complexity of the profiles, and deep spectro-polarimetry to increase signal-to-noise ratio of the observations.

\section{Summary and conclusions \label{sec:concl}}
In this paper, we have explored the diagnostic value of the \hed line observed with SST/CRISP during a C3.6-class flare. The observations showed both strong emission at the flare footpoints as well as strong absorption in the flare loops. There is polarimetric signal in all Stokes parameters in the flare footpoints and to a limited degree also in the flare loops. 

We demonstrate the temporal evolution of the flare footpoints in \hed, exhibiting downflows of up to $60\,\rm km~s^{-1}$ at the leading edge of the flare footpoints. All profiles show similar time-evolution, and are suggestive of strong plasma flows impacting and shocking the deep chromosphere, generating weak upflows as a result. \hed seems thusfar to be the only spectral line capable of so clearly capturing the effect of strong condensations on the deep chromosphere.

Chromospheric magnetic field and line-of-sight velocity maps for the flare are derived. The flare footpoints harbour very large values for the magnetic field of up to $\sim 2500$ G and host the strongest downflows, while the flare loops have a magnetic field between $0-500$ G and are stationary to modestly upflowing. Strong downflows are also found near the reconnection site. By combining all information obtained from the observations, we propose the magnetic field for the flare to have a fan-spine topology. Such inversion results are valuable to cross-validate with - or to use as input for - magnetic field extrapolations in flares.

It is clear from our observations that the \hed is very sensitive to the different physical regimes present in the flare. Therefore, we think that ground-based chromospheric flare observations should by default include \hed (or He~\textsc{i} 10830 \AA) if technically possible. \hed provides detailed information on dynamics and the magnetic field in the chromosphere during flares. Modeling of the lines could provide strong clues on energy transport mechanisms and electron acceleration models in flares. In the future, observations with increased polarimetric signal-to-noise ratio - such as will be provided by DKIST - will enhance the diagnostic potential of the lines even more.

\begin{acknowledgements}
The current work benefited from discussions with visiting researchers at the institute for Solar Physics in Stockholm. The authors specially wish to thank Carlos D\'iaz Baso, Andr\'es Asensio Ramos, Marian Mart\'inez Gonzalez, Lindsey Fletcher and Paulo Sim\~oes for useful advice on the presented work. We also wish to thank the anonymous referee for valuable comments which improved the content of the paper. We made use of routines by Rob Rutten to align SDO data with SST data. The observers thank Pit Sutterlin for support during the observing campaign at the SST.

TL acknowledges financial support from the
\textsc{Chromobs} project funded by the Knut and Alice Wallenberg
Foundation. JdlCR is supported by grants from
the Swedish Research Council (2015-03994) and the Swedish National Space Board
(128/15) and has received funding from the European Research Council (ERC) under the Euro-
pean Union’s Horizon 2020 research and innovation program (SUNMAG,
grant agreement 759548). JdlCR and SD are also supported by the Swedish Civil Contingencies Agency (MSB).
The inversions were performed on resources provided by the Swedish National Infrastruc-
ture for Computing (SNIC) at the High Performance Computing Center North
at Ume\aa~University.

The Swedish 1-m Solar Telescope is operated on the island of La Palma
by the Institute for Solar Physics of Stockholm University in the
Spanish Observatorio del Roque de los Muchachos of the Instituto de
Astrof\'isica de Canarias. 

This research has made use of NASA's
Astrophysics Data System Bibliographic Services.

\end{acknowledgements}

\bibliography{bibliography}

\begin{appendix}

\section{Attempts to increase signal-to-noise ratio \label{app1}}
We attempted to reduce the data without the application of MOMFBD since the effect of MOMFBD renders the distribution of the photon noise non-Gaussian and can in some cases lead to surpression of very faint signals. However, this reduction strategy introduced artifacts in the polarimetric calibration so we restrained from it and used the standard CRISPRED pipeline including MOMFBD (see Sect.~\ref{subs:sst}).

\begin{figure*}
\begin{center}
\includegraphics[scale=1]{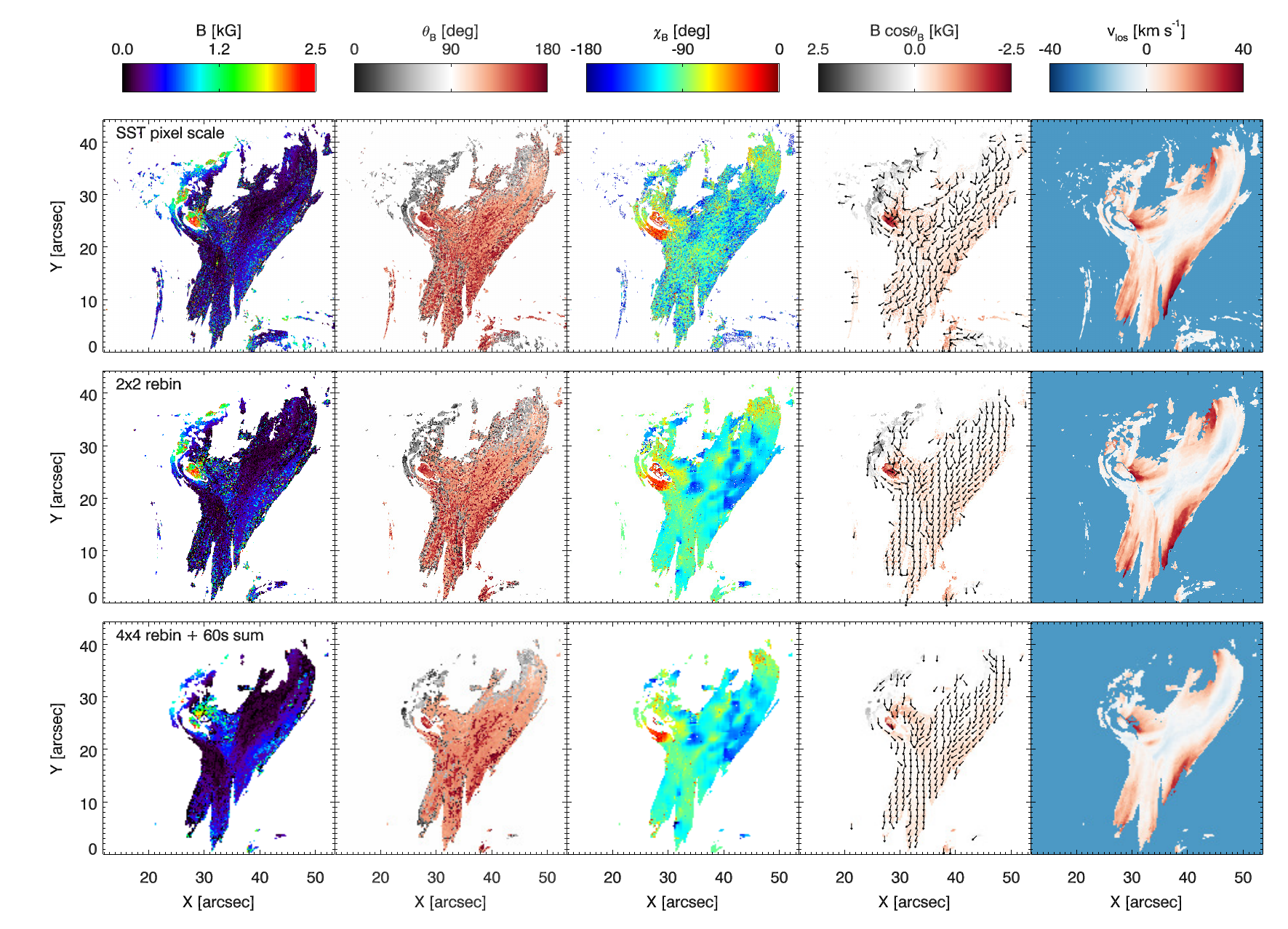}
\caption{Comparison of inversion results between the observation at SST pixel scale (top row), $2\times 2$ rebinning (middle row) and $4\times 4$ rebinning plus time-averaging of four time steps (bottom row). The columns show respectively the magnetic field strength $B$, the magnetic field inclination $\theta_B$, the magnetic field azimuth $\chi_B$ (input value determined by fibril orientation), the vertical magnetic field $B\cdot \cos\theta_B$ and the line-of-sight velocity $v_{\rm LOS}$. \label{fig:rebin}} 
\end{center}
\end{figure*}

We have also conducted several experiments with rebinning and time-averaging. We have run inversions on the data with the original SST pixel scale, on the data after spatial binning of $2\times 2$ pixels and on the data after spatial binning of $4\times 4$ pixels in combination with time-averaging of four time-steps (60 s). The results are shown in Fig.~\ref{fig:rebin}. Here, we do not focus on the scientific content of the maps but on the noise and the difference between binned and non-binned cases. As expected, the case with $4\times 4$ pixel rebinning and time-averaging resulted in the cleanest maps. However, even the map at the original SST pixel scale is acceptably clean and shows the same large-scale variation as the rebinned cases. The fact that the retrieved physical parameters are very similar in all three cases encourages our trust in the inversion results obtained with \textsc{Hazel}. It might be somewhat surprising that time-averaging does not introduce larger differences in the obtained maps as compared to single time-steps. However, the \hed absorption structure does not exhibit quick temporal evolution (see Sect.~\ref{sec:lineform}). Eventually, we decided to apply inversions to the data with $2 \times 2$ rebinning because this seemed the best compromise between reducing the inversion noise and preserving spatial information.  

\section{Influence of atomic orientation in the emission profiles\label{par:ator}}
Before we decided to introduce a second slab with different $v_{\rm LOS}$ and $B$ to fit the emission profiles, we have checked whether the strength and asymmetry of the observed Stokes $V$ profiles could not be modeled with atomic orientation instead. By default, the \textsc{Hazel} code includes the physics of atomic alignment and atomic alignment-to-orientation. Additionally, atomic orientation can be present in an observation when the incident light on the helium atoms is circularly polarized, or when the magnetically splitted components $\sigma_+$ and $\sigma_-$ are differentially illuminated. Both scenarios need a velocity difference between the photosphere generating the incident light and the layer where the scattering helium atoms are located (see e.g., \citealtads{2004ASSL..307.....L,2012ApJ...759...16M}). In the former case, the atomic orientation in \textsc{Hazel} is described by the parameter \begin{equation}
\frac{\bar{J}_0^1}{\bar{J}_0^0}=\sqrt{\frac{3}{2}}\frac{\int_0^\infty\int_0^1 V(\nu,\mu)\,\mu\, \der\mu\,\phi(\nu)\, \der \nu}{\int_0^\infty\int_0^1 I(\nu,\mu)\,\der\mu\,\phi(\nu)\,\der\nu},
\end{equation}  \citepads{2012ApJ...759...16M} and is by default set to zero, but can be given as an input parameter by the user. In theory, atomic orientation only equals the above expression for a two-level atom with an unpolarized lower level, and this assumption is invalid for the \hed line. However, our goal is to test the influence of atomic orientation on the strength and asymmetry of the Stokes $V$ profile, so we choose to operate under this incorrect assumption for testing purposes only. 

\begin{figure*}
\includegraphics[scale=1]{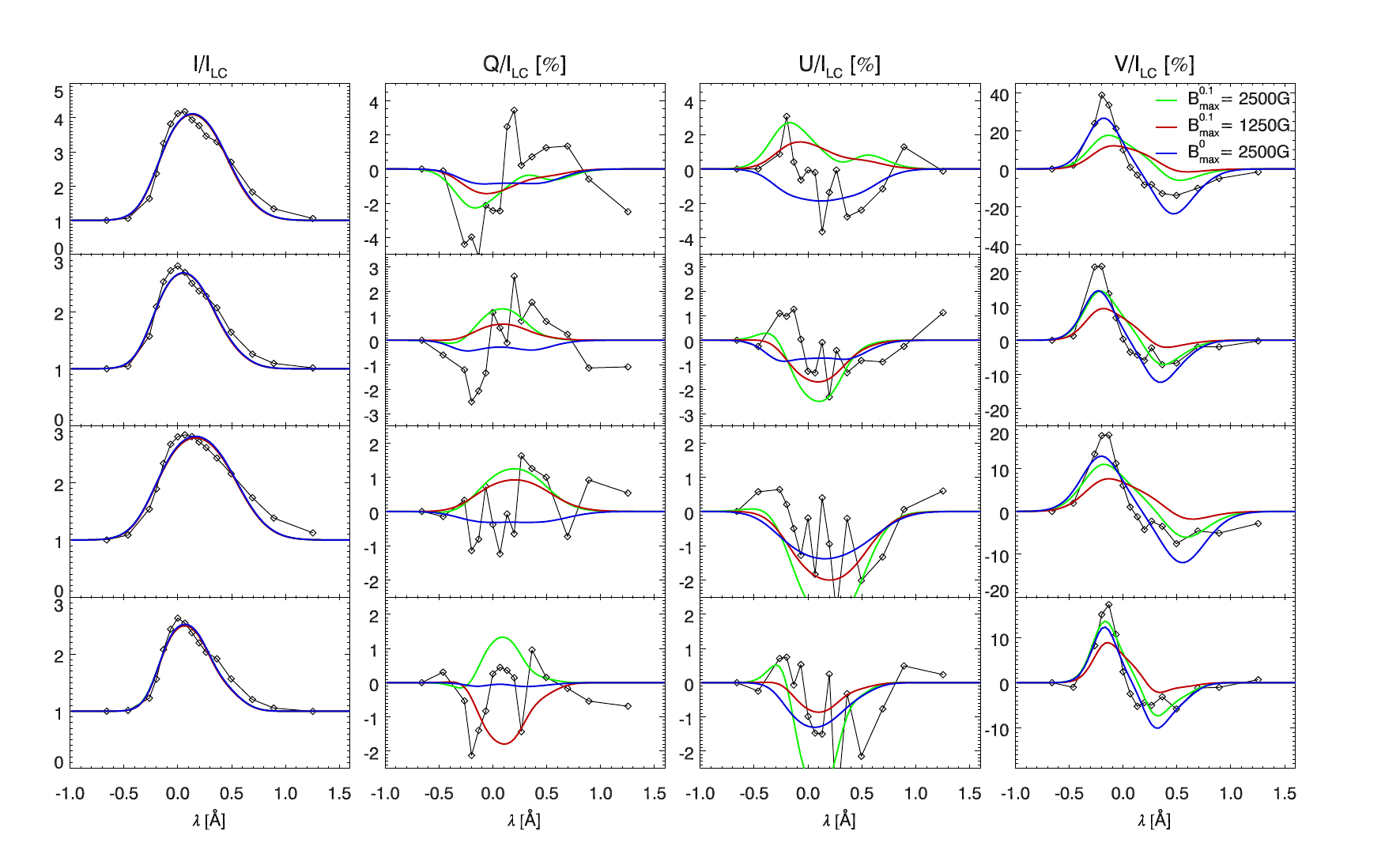}
\caption{Inversion results for the test cases shown in boldface in Table~\ref{testc} for a model using one slab in \textsc{Hazel}. The observational data is shown as a black line. Diamonds show the wavelength sampling of the observation. The green, red, and blue lines correspond to the fits obtained using the different upper values for the magnetic field $B_{\rm max}$ and different values for the atomic orientation defined by $\frac{\bar{J^1_0}}{J^0_0}$. The value for atomic orientation is given by a superscript to $B_{\rm max}$. All profiles are normalized using the local continuum $I_{LC}$.\label{ao1c}}
\end{figure*}

\begin{figure*}
\includegraphics[scale=1]{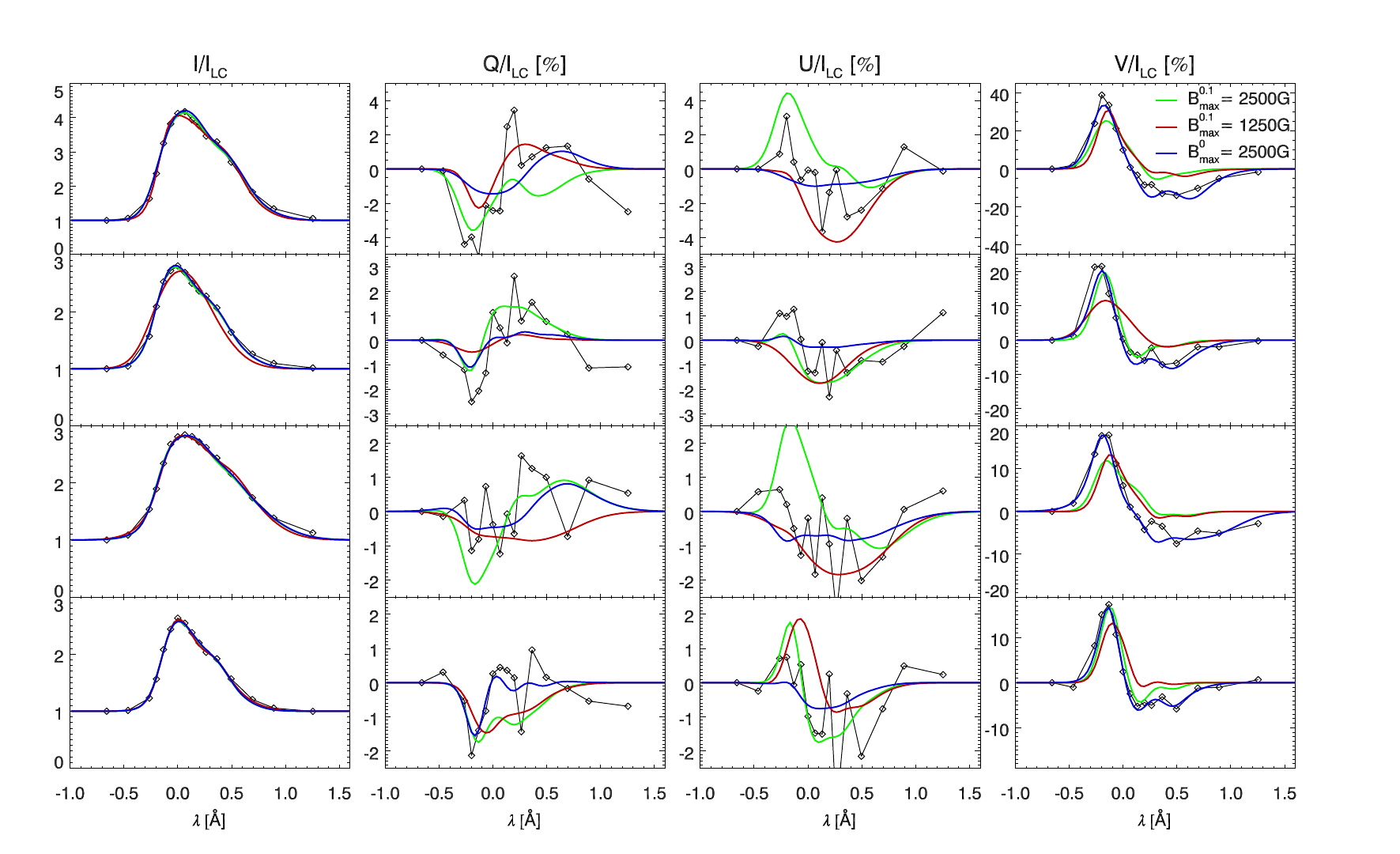}
\caption{Inversion results for the test cases shown in boldface in Table~\ref{testc} for a model using two slabs in \textsc{Hazel}. The observational data is shown as a black line. Diamonds show the wavelength sampling of the observation. The green, red, and blue lines correspond to the fits obtained using the different upper values for the magnetic field $B_{\rm max}$ and different values for the atomic orientation defined by $\frac{\bar{J^1_0}}{J^0_0}$. The value for atomic orientation is given by a superscript to $B_{\rm max}$. All profiles are normalized using the local continuum $I_{LC}$. \label{ao2c}}
\end{figure*}

We have run a series of test inversions on four selected profiles that have a strong and asymmetric Stokes $V$ profile, see Figs.~\ref{ao1c} and \ref{ao2c}. For each inversion, we have varied the value of $\frac{\bar{J}_0^1}{\bar{J}_0^0}$ for a model using one slab (Fig.~\ref{ao1c}) and a model using two slabs (Fig.~\ref{ao2c}). Table~\ref{testc} shows the chosen values for different test cases, each of those tested for both the one and two slab cases. We have checked that the chosen parameter values cover the necessary parameter space by synthesizing the corresponding profiles (not shown). A value of 0.001 for atomic orientation almost does not influence the \hed profile, while a value of 0.5 results in an almost completely postive Stokes $V$ profile, dominated by atomic orientation and not by the Zeeman effect. Our goal for these test runs was to answer the following questions: a) do we need one or two slabs to properly fit the Stokes $I$ and Stokes $V$ profiles? b) Is the asymmetry in the Stokes $V$ profile caused or influenced by atomic orientation? c) Can we obtain a strong and asymmetric Stokes $V$ profile with a lower value for the magnetic field and a higher value for atomic orientation?

The answer to question a) is that we need two slabs. This is clear from the comparison between the fit of the Stokes $I$ and Stokes $V$ component in Fig.~\ref{ao1c} and Fig.~\ref{ao2c}. We need two velocity components to fit the Stokes $I$ profiles and the asymmetry of the Stokes $V$ profile. The Stokes $Q$ and $U$ profiles are noisy and complicated and not fitted accurately in our inversions. The best we can say is that for the two slab case, the fits might be consistent with the profiles within the noise limit.

As to question b), it is clear that atomic orientation will not help the profile to reach the strong asymmetry that we observe. The large values for atomic orientation introduce an asymmetry but at the same time broadens the profile. The effect is that the Stokes $V$ profile becomes almost entirely positive and has a width comparable to the width of the Stokes $I$ profile. However, our observed asymmetry seems to maintain the width of the single lobes of Stokes $V$ but just enhances the intensity of the positive lobe and decreases the intensity of the negative lobe. This asymmetry is largely reproduced by the introduction of a second velocity component along the LOS. There is no need for atomic orientation to fit the Stokes $V$ profile. 

The answer to question c) is that we need those high values for the magnetic field in order to fit the strongest Stokes $V$ profiles, high values for atomic orientation cannot compensate for lower values of the magnetic field.

\begin{center}
\begin{threeparttable}

\caption{Parameter values for the series of test inversions. \label{testc}}
\begin{tabular}{p{2cm} p{2cm}}
$\frac{\bar{J^1_0}}{\bar{J^0_0}}$ & $B_{\rm max}$ [G]\\\hline
\bf{0} & \bf{2500}\\
0 & 1250 \\
0.001 & 2500 \\
0.01 & 2500 \\
\bf{0.1} & \bf{2500} \\
\bf{0.1} & \bf{1250} \\
0.5 & 2500 \\
0.5 & 1250\\
\end{tabular}

 \begin{tablenotes}
      \small
      \item  {\bf Note}. We tested all of these cases both for a model using one slab and a model using two slabs. The values in boldface corresponds to the test cases that are shown in Figs.~\ref{ao1c} and \ref{ao2c}.
    \end{tablenotes}
    
\end{threeparttable}
\end{center}


\section{Sensitivity of magnetic field strength and inclination to azimuth \label{app3}}
In Sect.~\ref{sec:loop}, we describe how we used an estimate for the fibril orientation of the flare loops as an input for the value of the azimuth $\chi_B$ in the inversion. However as shown in the third column of Fig.~\ref{fig:inv}, the output of the inversion for the azimuth is almost the same as what we put in. As it turns out, the Stokes $Q$ and $U$ signals are too noisy for \textsc{Hazel} to prefer one value of the azimuth over another, and the code converges to values of what is very close to the input value. This means that we cannot trust the azimuth values that came out of the inversions. However, in the local vertical frame, the values of $B$ and $\theta_B$ are influenced by the value of the azimuth. In order to quantify how sensitive these parameters are to $\chi_B$, we have performed two different tests. 

\begin{figure*}
\includegraphics[scale=1]{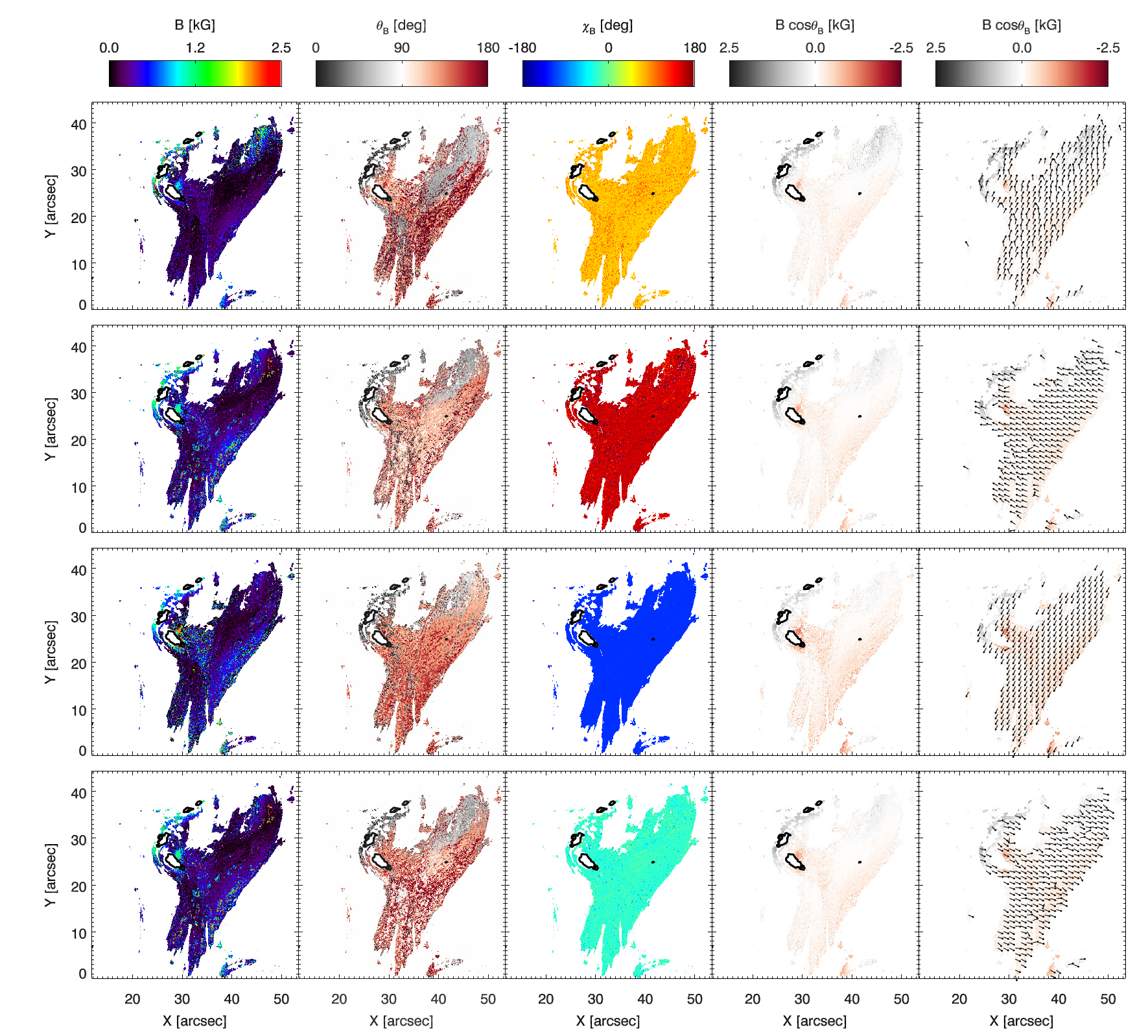}
\caption{Test of the sensitivity of $B$ and $\theta_B$ to different input values of $\chi_B$. Top to bottom corresponds to input values of $\chi_B=[-115,-25,65,155]\degree$. The test was only performed on the absorption profiles. The black contours indicate the locations of the emission profiles. \label{testaz}}
\end{figure*}

The result of the first test is shown in Fig.~\ref{testaz} and shows how the maps are affected when using different input values for the azimuth. We have tested four different input values: $\chi_B=[-115,-25,65,155]\degree$ that were constant over the field-of-view. We let \textsc{Hazel} find the best fit using the Levenberg-Marquardt algorithm for the azimuth and the DIRECT + Levenberg-Marquardt algorithm for the other parameters. We conclude that the output azimuth is nearly equal to the input azimuth in almost all cases. {However, the map for $B$ does not exhibit extreme qualitative changes: they are very similar in all four cases}, with some generally lower values in the case of $\chi_B=65\degree$. The maps for the inclination have similar polarities in all four cases, {although the regions of positive polarity do show variations in size. The inclination in the bulk of the flare also shows changes of more vertical toward more horizontal field. Therefore, we should be cautious with the interpretation of the inclination.} The case of $\chi_B=-25\degree$ is on average the most similar to our result that we presented in Fig.~\ref{fig:inv}, where we have used using fibril orientation as azimuth input. We see that in the case of $\chi_B=-25\degree$, both the maps of the magnetic field and of the inclination are quite smooth. This could be taken as a hint that our assumption that the field is mostly aligned with the direction of the fibrils could be valid.

\begin{figure}
\includegraphics[scale=1]{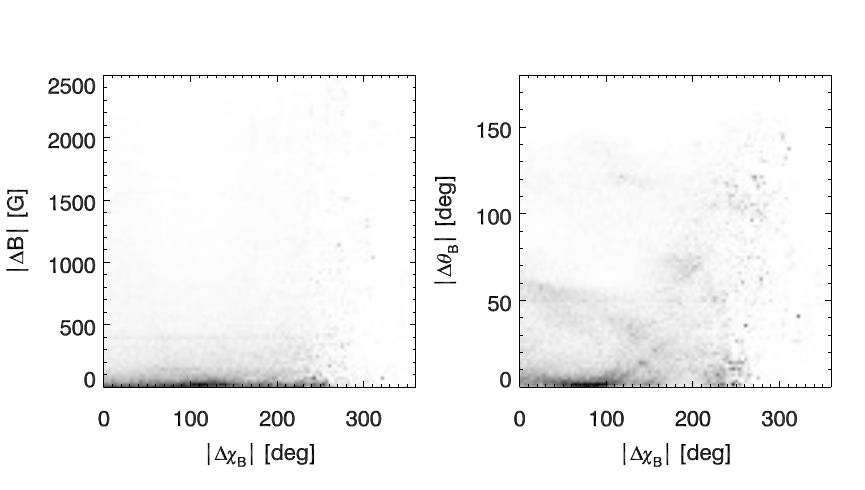}
\caption{Variation of the magnetic field difference  $|\Delta B|=|B_{\rm step}-B_{\rm loop}|$ and the inclination difference $|\Delta \theta_{B}|=|\theta_{B,\rm step}-\theta_{B,\rm loop}|$ with the azimuth difference $|\Delta \chi_B|=|\chi_{B,\rm step}-\chi_{B,\rm loop}|$. Each column is normalized with its total number of elements. \label{sens}}
\end{figure}

For the second test, we have randomly selected $10\%$ of the absorption pixels at $t_{\rm peak}$ and varied the azimuth in steps of $10\degree$ between $-180\degree$ and $180\degree$. We called this input azimuth $\chi_{B,\rm step}$. Proceedingly, we used this value $\chi_{B,\rm step}$ as input in the inversion in the same way as described above. We compare the results for the magnetic field $B_{\rm step}$ and $\theta_{B,\rm step}$ to the original result obtained by using the loop orientation estimate as input: $B_{\rm loop}$ and $\theta_{B,\rm loop}$. In Fig.~\ref{sens}, we show the variation of the magnetic field difference $|\Delta B|=|B_{\rm step}-B_{\rm loop}|$ and the inclination difference $|\Delta \theta_{B}|=|\theta_{B,\rm step}-\theta_{B,\rm loop}|$ with the azimuth difference $|\Delta \chi_B|=|\chi_{B,\rm step}-\chi_{B,\rm loop}|$. The magnetic field strength turns out to be very stable and almost insensitive to the azimuth changes. The inclination is a bit more sensitive, as expected, especially for large differences between the azimuth values but the largest population is still found at around zero sensitivity for the angles between $0\degree$ to $90 \degree$. We conclude that Figs.~\ref{testaz} and \ref{sens} demonstrate the robustness of our results since the magnetic field strength $B$ and the inlination $\theta_B$ do not heavily depend on our assumption of the magnetic field being aligned with the fibrils.

\end{appendix}

\end{document}